# Generation of the Cluster States using Double Quantum Dots in Circuit QED

Mir Massoud Aghili Yajadda*
School of Physics
University of Sydney

(14[th] December 2009)

*Electronic Address: **maghili@physics.usyd.edu.au**

# Abstract

The cluster state quantum computation is a versatile approach to build a scalable quantum computer. In this thesis we theoretically demonstrate that a one dimensional array of double quantum dots with long spin relaxation time can evolve to a cluster state via capacitive coupling to a transmission line resonator. By applying an oscillating voltage to gates of the double quantum dots, we prove that the evolution operator for such interaction is an Ising-like operator. Subsequently, the interacting qubits will become highly entangled that we realize a cluster state. Furthermore, we propose an experiment to investigate validity of our theoretical method. Considering the current advanced technology in semiconductor nanofabrication, our proposed structure can be integrated on a chip where provides scalability and convenient control of the qubits. The scalability of our scheme implies that we can expand this structure to higher dimensional arrays of the qubits where paves the way for further experimental investigation on the theory of measurement-based quantum computation.

# Contents

# 1. Literature Review

## 1.1 Introduction

Information technology has come a long way since building the first computer decades ago and the importance of computers in our daily life is almost undeniable; the current technologies in manufacturing semiconductor devices has made substantial improvements to the speed, reliability and portability of processors by reducing the size of electronic components to nano-scales as well as improving the quality of materials. However, the laws of quantum mechanics impose fundamental limits on the minimum size of electronic devices. Therefore, in order to improve on current technologies, quantum mechanical devices need to be developed.

In classical computers, computations are performed by applying logical operations to classical bits, which can be in one of two states, 0 or 1. In modern computers, this is implemented at the nano-scale, where the quantum mechanical characteristics of the system become more important. In order to study the behavior of such devices we are required to apply the laws of quantum mechanics. One of the characteristics of quantum mechanics is the quantization of energy levels; hence we may consider the excited state and the ground state of a quantum system to represent a quantum bit (qubit). These energy states could be charge states or electrons spin states depending on the quantum structures. However, while such devices are quantum mechanical systems, the computation is still completely classical, since the method of processing information is based upon physical devices that implement classical logic gates.

In contrary to classical computation qubits can display quantum mechanical properties where the power of Quantum Computation (QC) may be limited only by quantum noise. Quantum computation provides new resources that cannot be utilized in classical computation. Firstly, qubits can exist in superposition of states. Consider a quantum system that can be in one of the quantum states $\Psi_A$ or $\Psi_B$, which are both elements of some Hilbert space. Then, according to the laws of quantum mechanics, any complex superposition of $\Psi_A$ and $\Psi_B$ is also a valid state that the system could be in. In Quantum Information Processing (QIP) we employ quantum gates to perform QC. Unlike classical logic gates which are physical objects (such as transistors and diodes), a quantum gate is not an electronic component. Rather, a quantum gate is abstract transformation acting on a quantum state. One way of implementing a quantum gate is to cause a quantum system to interact with surrounding fields in a prescribed manner. The evolution operator of the quantum state then acts as a quantum gate. However, only a limited amount of quantum gates have been reliably implemented, so a preliminary step towards QC is to develop the ability to reliably implement a variety of different quantum gates.

A second resource available for QC is *entanglement*. Entanglement is a remarkable quantum mechanical phenomenon that can be used to achieve QC. The following example demonstrates the concept of entanglement. Consider two electrons which are in the singlet state of an atom. According to Pauli's exclusion principle two electrons with similar spin orientation cannot occupy the same energy level hence the spin orientation of these electrons will be in opposite direction (singlet state). Now we take these electrons apart from each other and we measure one of the electrons in any basis. The measurement outcome then tells us not only the state of the electron that was measured, but also the state of the other electron at the same time. This demonstrates that electrons were in an entangled state and by performing the measurement we have destroyed the entanglement as further measurements on one electron will provide no new information about the other electron.

One quantum system that could allow for efficient QC is the *cluster state.* In an N-dimensional lattice where qubits occupy different lattice sites the interaction of qubits with their neighbors can generate a highly entangled state known as the cluster state. A cluster state can be generated by introducing an interaction Hamiltonian between arrays of qubits.

Since measurement on a qubit at one site effects the quantum states of entangled qubits, performing measurements on single qubits in cluster states can produce quantum circuits, which implement a desired quantum gate on an input state. This approach to creating quantum circuits by using measurements is called *measurement-based QC*. One of the interesting features of the cluster state QC is the *one-way* QC; as measurement on qubits destroys the qubit entanglement we can use the cluster states only once which makes QC one-way.

In our proposal to generate a cluster state we employ semiconductor Quantum Dot (QD) to represent a qubit using a two level system. Quantum dots are quantum systems where electrons are spatially confined in all dimensions (giving a zero dimension), resulting in discrete energy levels. Electron's long spin coherence time in the semiconductor QD makes this system a promising candidate for QIP; furthermore due to the low electron density in the QD, electrons will have large Fermi wave lengths. Therefore, we can conveniently manipulate and control the electronic properties of a QD by applying an external electromagnetic field. QDs also provide a suitable bridge between classical and quantum computers, as they can be implemented using semiconductor nanotechnology, which can be integrated with current electronic devices. But we should bear in mind that there will be technical challenges in the fabrication of QDs, for example size variation and defect formation. In this thesis our emphasis will be on the semiconductor QDs because of the abovementioned promising features.

In this thesis we demonstrate that we can theoretically generate the cluster states from 1D array of interacting qubits. The evolution operator is an Ising-like operator where the quantum system evolves to the clusters in presence of the noise related fluctuations. The implementation of the structure is compatible with current nanotechnology, allowing it to be integrated on a chip. This provides a convenient of addressing and controlling qubits. In addition to these features, the cluster states are created in one step which makes this scheme a promising candidate to build a scalable quantum computer.

Our discussion in section 1.2 is a brief overview of QIT then it is followed by definition of the relaxation and dephasing times. Since the atom-cavity interaction is the fundamental part of this thesis, in section 1.4 a standard Hamiltonian is introduced to analyze such interactions. Afterward, a superconducting microstrip line resonator is represented that acts as a cavity to supply microwave photons. Later, the quantum systems with properties similar to real atoms are discussed which can be used as qubits. At the end of this chapter some information about our proposed qubit and its energy structure is provided.

In chapter two, we investigate a generalization of a theory which has been demonstrated in ion trap experiment. This model is the theoretical basis of this thesis, and so we explore it in some detail that we investigate this model in chapter two. Then we extend this method to our proposal in chapter three where we theoretically demonstrate that the cluster states can be generated in our scheme. In chapter four we propose an experiment to perform the single-qubit readout, which is part of the measurement based quantum computation.

## 1.2 Basics of the quantum information theory

This section is a brief introduction to the basics of quantum computation, a more comprehensive overview of the subject can be found in reference [1].

The Quantum bit is the fundamental concept of QIT and it is the quantum complement of the bit in the classical computation. The qubit states are a complex vector space and can be represented in the computational basis by normalized complex superposition of the two computational basis states, $|0\rangle$ and $|1\rangle$. The single-qubit state $|\psi\rangle=\cos(\theta/2)|0\rangle+e^{i\varphi}\sin(\theta/2)|1\rangle$ can be displayed using Bloch sphere (figure 1.1) and this sphere has unit radius as qubit states are normalized [1].

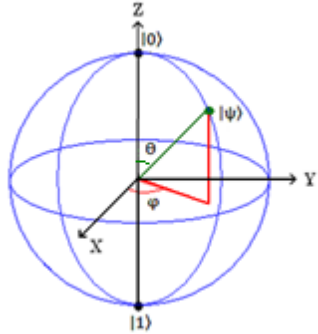

*Figure 1.1: Demonstration of a single-qubit state in Bloch sphere.*

One method of implementing QC is applying a sequence of qubit gate operations to a set of qubits. Qubit gate operations are unitary operators that act on the qubits states. These gates can be single or multiple qubit gates. Quantum gates and qubit states can be represented by matrix formalism. For example, the computational basis states and the single qubit Hadamard gate (Ha) can be represented by [1]

$$|0\rangle = \begin{pmatrix}1\\0\end{pmatrix} \qquad |1\rangle = \begin{pmatrix}0\\1\end{pmatrix} \qquad (Ha) = 1/\sqrt{2}\begin{pmatrix}1 & 1\\1 & -1\end{pmatrix} \qquad (1.1)$$

$$(Ha)|0\rangle = 1/\sqrt{2}(|0\rangle + |1\rangle) \qquad (Ha)|1\rangle = 1/\sqrt{2}(|0\rangle - |1\rangle) \qquad (1.2)$$

An important quantum gate is the controlled-NOT gate (CNOT). The CNOT gate acts on two qubits, a control and a target qubit. The two qubit input states can be represented as a complex superposition of the four computational basis states for two qubits. The four computational basis states for two qubits and the CNOT gate can be represented in matrix form by [1]

$$|00\rangle = \begin{pmatrix}1\\0\\0\\0\end{pmatrix} \quad |01\rangle = \begin{pmatrix}0\\1\\0\\0\end{pmatrix} \quad |10\rangle = \begin{pmatrix}0\\0\\1\\0\end{pmatrix} \quad |11\rangle = \begin{pmatrix}0\\0\\0\\1\end{pmatrix} \quad CNOT = \begin{pmatrix}1 & 0 & 0 & 0\\0 & 1 & 0 & 0\\0 & 0 & 0 & 1\\0 & 0 & 1 & 0\end{pmatrix} \qquad (1.3)$$

$i$ & $j$ in a two qubit state $|ij\rangle$ represent the control qubit and the target qubit respectively. From the matrix form, the CNOT gate can be seen to act on kets according to [1]

CNOT $|00\rangle = |00\rangle$   CNOT $|01\rangle = |01\rangle$   CNOT $|10\rangle = |11\rangle$   CNOT $|11\rangle = |10\rangle$ (1.4)

Since the CNOT gate can be expressed as a product of two Ha gates and a controlled-Phase (C-Phase) gate, [1]

$$\text{CNOT} = \mathbf{Ha}\text{ C-Phase }\mathbf{Ha} \qquad \mathbf{Ha} \equiv \begin{pmatrix} \text{Ha} & 0 \\ 0 & \text{Ha} \end{pmatrix} \qquad \text{C-Phase} = \begin{pmatrix} 1 & 0 & 0 & 0 \\ 0 & 1 & 0 & 0 \\ 0 & 0 & 1 & 0 \\ 0 & 0 & 0 & -1 \end{pmatrix} \qquad (1.5)$$

a CNOT gate can be experimentally realized using C-Phase and Ha gates. CNOT gates are important because an arbitrary quantum gate can be implemented with arbitrary accuracy using only single qubit and CNOT gates (i.e. single qubit gates and the CNOT gate form a universal set of quantum gates). Since CNOT gates can be implemented exactly using Ha gates (which are single qubit gates) and a C-Phase gate, it follows that single qubit gates and the C-Phase gate also form a universal set of quantum gates [1]. This important feature of the CNOT (and equivalently, the C-Phase) gate makes them one of the most useful gates for quantum computation.

### 1.2.1 The cluster states

Cluster states are sometimes referred as *graph states*, [2] since they can be represented using graphs. Figure 1.2 displays a 1D array of qubits that evolve to a cluster state due to an Ising-like interaction, the cluster state representing this graph can be generated by preparing the qubits in the state $|+\rangle = (|0\rangle+|1\rangle)/\sqrt{2}$ where a C-Phase gate is applied between the qubits [2].

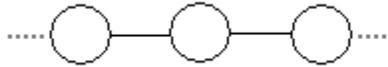

*Figure 1.2: Schematic of a cluster state using graph representation. Circles (graph vertices) represent qubits to which a C-Phase gate is applied.*

Measurement based QC is achieved by performing a series of single-qubit measurements (processing measurements). The measurement basis depends on the result of the preceding measurement. There are two alternative ways to define the result of the computation. We can consider the result as being a quantum state determined by the state of the qubits which have not been measured. Or we can perform the single-qubit readout measurements on each of these qubits and store the results as an array of logical bits in a classical computer [2].

The single-qubit measurement will destroy the entanglement, hence we can use cluster states as a substrate to imprint any quantum circuit [3]. This will allow us to transport the information along a certain direction (figure 1.3), so it is possible to create wires on the cluster states to send information along a specific path.

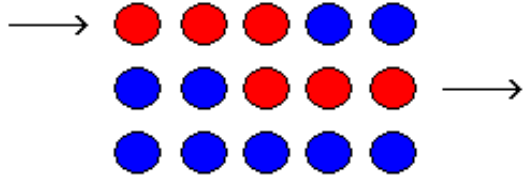

*Figure 1.3: Illustration of an imprinted wire on a cluster state. Arrows indicate the direction in which information flows. Circles represent qubits. The blue circles are the qubits that single-qubit measurement has been performed on them.*

## 1.3 Relaxation and Dephasing

The amount of time it takes for an atom to decay from the excited to the ground state due to spontaneous emission is called the relaxation time; this radiative decay can be estimated using Fermi's golden rule [4]. Although there is no clear theoretical explanation of spontaneous emission, it is thought that the sources of relaxation due to spontaneous emission depend on the atomic structures and the way in which atoms interact with any surrounding fields. From Fermi's golden rule, the transition rate for an atom due to interaction with a cavity field is [4]

$$\gamma = \frac{2\pi}{\hbar}\rho(\omega)(|\langle e, 0|H|1, g\rangle|^2 \delta(E - \hbar\omega) \qquad \hbar = h/2\pi \qquad (1.6)$$

where $h$ is Planck's constant, $\rho(\omega)$ is the energy spectrum of the cavity, H is the interaction Hamiltonian, $|1, g\rangle$ indicates that the atom is in the ground state with one photon in the cavity with energy of $\hbar\omega$, state $|0, e\rangle$ indicates the atom is in the excited state and no photons are in the in the cavity. The last term a delta function ensures that the energy of the photon is equal to the excitation energy E of the atom so the process will conserve energy. From this the relaxation time is T1= $1/\gamma$.

Another parameter for assessing the feasibility of an experiment is dephasing, which refers to the randomization of the phase difference between two qubit states. Dephasing can occur due to many different physical processes, depending on the physical systems being used. For example, it may originate from the 1/f noise due to abnormal behavior of the junctions, properties of the materials or the spin-orbit interaction. We estimate the dephasing time in GaAs double quantum dot by [5]

$$T2^* \sim \hbar/(g^* \mu_B B_{nuc}) \qquad (1.7)$$

$T2^*$ is the ensemble spin dephasing time, $B_{nuc}$ is the nuclear magnetic field, $\mu_B$ is the Bohr magneton and $g^*$ is the effective electron g factor. Dephasing also can be visualized by projecting two qubit states on the x-y plane in Bloch sphere (figure 1.1) and recognizing the randomization of the phase difference between two qubits as the dephasing.

## 1.4 Jaynes-Cumming Hamiltonian

One physical implementation of a qubit is a two level atom where the ground state of the atom can be represented by $|0\rangle$ and its excited state by $|1\rangle$. Therefore, QIP will be achievable by encoding qubit states in the energy eigenstates of a two level atom. While no two level atoms exist in the physical world, they can be approximated in a cavity by selecting an atom for which there is only one transition whose transition energy E matches the lowest energy of the cavity photon modes. The interaction Hamiltonian that describes the evolution of these structures is the key for analyzing quantum gate operations and has been studied broadly in quantum optics [6].

We will investigate the interaction between a single mode classical field and a two level atom by appealing to a semi-classical technique. To do so, we can write the wave function of a two level atom as [6]

$$|\psi(t)\rangle = a_g(t)|g\rangle + b_e(t)|e\rangle \qquad |a_g(t)|^2 + |b_e(t)|^2 = 1 \tag{1.8}$$

$|g\rangle$ & $|e\rangle$ are the ground state and the excited state of a two level atom with energy eigenvalues $\hbar\omega_g$ & $\hbar\omega_e$ and $|a_g|^2$ & $|b_e|^2$ are the probability of being in the ground state or in the excited state respectively. The evolution of the atom is described by Schrödinger's equation [6]

$$\frac{d|\psi(t)\rangle}{dt} = -i\,\hat{H}|\psi(t)\rangle \qquad \hat{H} = \hat{H}_0 + \hat{H}_{int} \tag{1.9}$$

$\hat{H}_0$ and $\hat{H}_{int}$ are the unperturbed and the interaction Hamiltonian respectively which can be written as [6]

$$\hat{H}_0 = \hbar\omega_g\,|g\rangle\langle g| + \hbar\omega_e\,|e\rangle\langle e| \tag{1.10}$$

$$\hat{H}_{int} = -erE(t) = -(d_{ge}|g\rangle\langle e| + d_{eg}|e\rangle\langle g|)E(t) \tag{1.11}$$

$E(t)$ represents the electric field at the position of the atom and $d_{ge} = d^*_{eg} = e\langle g|r|e\rangle$ forms the matrix element of the electric dipole moment. In the presence of a linearly polarized electric field and considering the dipole approximation (assuming the uniform electric field at the position of the atom) we can obtain the time variations of the probability amplitudes to be [6]

$$a_g(t) = \left\{ a_g(0)\left[\cos\left(\frac{vt}{2}\right) + \frac{i\delta}{v}\sin\left(\frac{vt}{2}\right)\right] + \frac{i\Omega}{v}e^{i\varphi}a_e(0)\sin\left(\frac{vt}{2}\right)\right\}e^{\frac{-i\delta t}{2}} \tag{1.12}$$

$$a_e(t) = \left\{ a_e(0)\left[\cos\left(\frac{vt}{2}\right) - \frac{i\delta}{v}\sin\left(\frac{vt}{2}\right)\right] + \frac{i\Omega}{v}e^{-i\varphi}a_g(0)\sin\left(\frac{vt}{2}\right)\right\}e^{\frac{i\delta t}{2}} \tag{1.13}$$

$$E(t) = E_0\cos(ft) \qquad \Omega = \frac{|d_{eg}|E_0}{\hbar} \qquad d_{eg} = |d_{eg}|e^{i\varphi} \tag{1.14}$$

$$\omega = \omega_e - \omega_g \qquad \delta = \omega - f \qquad v = \sqrt{\Omega^2 + \delta^2} \tag{1.15}$$

Here $E_0$ & $f$ are amplitude and frequency of the electric field, $\Omega$ represents the Rabi frequency, $\omega$ is the transition frequency of the atom and phase of the electric dipole matrix element is denoted by $\varphi$. In obtaining (1.12) and (1.13) we need to ignore the high frequency term ($\omega + f$) (referred as the Rotating Wave Approximation (RWA)). This is reasonable approximation that we will use repeatedly while studying atom-field interaction where we neglect energy violating processes [6].

Assuming that the atom is initially at the ground state, then the maximum coupling between the field and the atom will take place at the resonance condition ($\delta = 0$) and the population inversion ($|a_g(t)|^2 - |b_e(t)|^2$) will oscillate in time with Rabi frequency [6]. This phenomenon is essential for QC; in the next chapter we will see how a quantum gate is achieved when Rabi frequency satisfies a certain condition.

Now look at the interaction between a single mode quantized cavity field and a two level atom, the total Hamiltonian (H) of this system is written as [6]

$$\hat{H} = \hat{H}_{cavity} + \hat{H}_0 + \hat{H}_{int} \qquad \hat{H}_{cavity} = \hbar f \left( \hat{a}^\dagger \hat{a} + \frac{1}{2} \right) \qquad (1.16)$$

$$\vec{E} = \vec{E}_0(\hat{a} + \hat{a}^\dagger) \qquad |\vec{E}_0| = \sqrt{\frac{\hbar f}{2\varepsilon_0 V}} \qquad (1.17)$$

where V is the volume, $\hat{a}^\dagger$ and $\hat{a}$ are the photon creation and annihilation operators respectively. Furthermore by using (1.14) and (1.15) we can rewrite H as [6]

$$\hat{H} = \hbar f \hat{a}^\dagger \hat{a} + (1/2)\hbar\omega\hat{\sigma}_z + \hbar g(\hat{\sigma}_+ + \hat{\sigma}_-)(\hat{a} + \hat{a}^\dagger) \qquad \hbar g = -\vec{d}_{eg} \cdot \vec{E}_0 \qquad (1.18)$$

$$\hat{\sigma}_z = |e\rangle\langle e| - |g\rangle\langle g| \qquad \hat{\sigma}_+ = |e\rangle\langle g| \qquad \hat{\sigma}_- = |g\rangle\langle e| \qquad (1.19)$$

Here we have excluded zero point energy from Hamiltonian of the cavity, $\hat{\sigma}_z$ is one of the Pauli spin matrices, $\hat{\sigma}_+ = 1/\sqrt{2}(\hat{\sigma}_x + i\hat{\sigma}_y)$, $\hat{\sigma}_- = 1/\sqrt{2}(\hat{\sigma}_x - i\hat{\sigma}_y)$ are written based on the other two Pauli spin matrices and $\hbar g$ is the coupling strength. Finally we can write the last term of the H as the Jaynes-Cumming (JC) Hamiltonian [6]

$$\hat{H}_{JC} = \hbar g(\hat{a}\hat{\sigma}_+ + \hat{a}^\dagger\hat{\sigma}_-) \qquad (1.20)$$

In obtaining this Hamiltonian we have omitted non energy conserving terms in the RWA and dropped a constant energy term $1/2(\hbar\omega_g + \hbar\omega_e)$. Considering a cavity in resonance with the transition frequency of a two level atom, we can describe the excitation process of the atom (photon absorption) by the $\hat{a}\hat{\sigma}_+$ term and its relaxation from the excited state to the ground state (photon emission) by the $\hat{a}^\dagger\hat{\sigma}_-$ term. The importance and capability of JC Hamiltonian in studying atom-field interaction have been investigated in the previous schemes such as cavity QED [7], circuit QED [8] and ion trap experiments [9][10].

## 1.5 Transmission line resonator

Now that we have the proper mathematical tools to study the interaction between the cavity and the atoms, the next step will be to investigate the structure where atoms can be entangled by a cavity mediated coupling. This can be realized by a Transmission Line Resonator (TLR). Interested readers can find the detailed derivation of the TLR's voltage operator in appendix A.

A Transmission Line Resonator is a superconducting 1D microwave strip line resonator; we can produce microwave photons by applying an oscillating voltage (V) to the TLR. In order to study the interaction between a TLR and its neighboring atoms we need to derive the Hamiltonian of the TLR. First we treat the TLR as an infinite series of inductors with inductance per unit length $l$ which are each grounded through a capacitor with capacitance per unit length $C_0$ (Figure 1.4). Then we write the Lagrangian of the TLR as [7][11]

$$\mathcal{L} = \int_{L/2}^{L/2} \left[ \frac{l}{2} \left( \frac{\partial Q}{\partial t} \right)^2 - \frac{1}{2C_0} \left( \frac{\partial Q}{\partial x} \right)^2 \right] dx \tag{1.21}$$

Q(x,t) is the charge on the TLR at any instance of time and L is the length of the TLR. By using calculus of variations we can extract the Euler-Lagrange equation that describes a wave traveling along the TLR at a speed u. To solve the wave equation we treat charge conservation as a boundary condition [11].

The next step is the quantization of the electric field created by the TLR. To do so, we provide $\hat{a}_n^\dagger$ and $\hat{a}_n$ as the photon creation and annihilation operators respectively. Defining canonically conjugated momentum, we can obtain the voltage operator for an infinite TLR [11]

$$\hat{V}(x,t) = \frac{1}{c_0} \frac{\partial Q(x,t)}{\partial x} = -\sum_{n_o} \sqrt{\frac{\hbar \omega_{n_o}}{Lc_0}} \sin\left(\frac{n_o \pi x}{L}\right) [\hat{a}_{n_o}(t) + \hat{a}_{n_o}^\dagger(t)] +$$

$$\sum_{n_e} \sqrt{\frac{\hbar \omega_{n_e}}{Lc_0}} \cos\left(\frac{n_e \pi x}{L}\right) [\hat{a}_{n_e}(t) + \hat{a}_{n_e}^\dagger(t)] \tag{1.22}$$

$n_o$ and $n_e$ are the number of odd and even cavity modes respectively. Due to finite length of the TLR, it will act as a resonator with resonant frequency of $\omega_0 = \frac{2\pi}{L\sqrt{lc_0}}$.

In order to carry out any measurement on the TLR it needs to be coupled to the outside world by capacitance C that makes renormalization of the resonant frequency inevitable. By considering the effective permittivity ($\varepsilon_0 = \frac{C}{Lc_0}$) of two gaps at both ends of the TLR (figure 1.4) we find the renormalized frequency to be $\omega \approx \omega_0(1-2\varepsilon_0)$ [12].

Finally for the first even frequency of the fundamental mode of the TLR we can write the voltage operator as [12]

$$\hat{V}_{TLR}(x) = \sqrt{\frac{\hbar \omega}{Lc_0}} \cos\left(\frac{\pi x}{L} + \varphi\right)[\hat{a} + \hat{a}^\dagger] \tag{1.23}$$

where $\varphi$ is the phase shift due to renormalization of the resonant frequency of the cavity and satisfies $\tan\varphi = 2\pi\varepsilon_0$ [12].

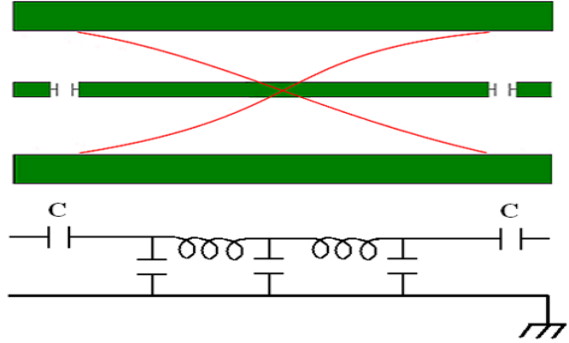

*Figure 1.4: Illustration of a TLR (green).The first even fundamental mode of the TLR (red). Equivalent circuit diagram of the TLR (black).*

## 1.6   The artificial atom

From studying the atom-field interaction in section 1.4 we recognize that the coupling strength $\hbar g$ is directly related to the strength of the atomic dipole moment (assuming constant electric field for the cavity in equation 1.18). To tackle the technical difficulties in dealing with real atoms and significance of the qubit gate operation speed, the first step is to build artificial atoms so that we can have control over atomic positions and transition frequencies as well as their dipole moment strength [13].

Similar to real atoms where electrons are confined in a potential well with quantized energy levels, we can artificially generate such electrostatic potential wells by using current advanced nanotechnologies in semiconductor nanofabrication. The current capability of building artificial atoms leads us to the next step, finding a way to generate entangled states from these qubits. Taking into account the transition frequency of these qubits (which is in the microwave frequency spectrum) we appeal to the techniques developed in the radio frequency (RF) industry.

Involvement in such technologies takes us to a new chapter in quantum information technology which is called circuit Quantum Electrodynamics (QED). For the first time the realization of the qubit entanglement in circuit QED was demonstrated in an on-chip two qubit quantum processor. This experiment was done by employing a microwave strip line resonator which acted as a cavity mediated coupling between two qubits [8].

## 1.7   The double quantum dot

In this section we introduce the structure of a double QD and we investigate the possibility of encoding qubits in the energy eigenstates of this system. Although there are some proposals to exploit GaAs double QD to perform the qubit gate operations, the analysis of such schemes is beyond the scope of this thesis. However, interested readers can obtain more information in reference [14] where one of the proposals for spin qubit manipulation has been investigated systematically.

## 1.7.1 Fabrication of a double QD

A quantum dot is an artificial atom that supplies electrons where they can be transferred in and out of the QD via tunnel coupling to the reservoirs (ohmic contacts which are connected to the voltage source). In this project we study a QD where both of the source and the drain are positioned at the same side of the structure. The GaAs QD which we analyze here is the product of the molecular beam epitaxy technique that is employed to fabricate QD from heterostructure of GaAs and AlGaAs. The AlGaAs region is doped with Si to create a thin layer (~10 nm) of free electrons at interface of the GaAs-AlGaAs (figure 1.5.a). This two dimensional electron gas (2DEG) is produced 50-100 nm beneath the surface with high mobility in the order of $10^5$-$10^7$ cm$^2$/Vs and low electron density (~1-5x$10^{15}$m$^{-2}$) [14].

The technology of the electron-beam lithography allows us to implement the metal gates on the surface of a QD in any shape we require. By applying a voltage to these gates we are able to control the number of electrons residing in each QD, where the number of electrons in each QD can be estimated by electrostatic charging energy of each QD. Electrons in a double QD can penetrate through the potential barrier by a tunneling process and the tunneling rate can be controlled by applying a voltage to a certain gate on the double QD (figure 1.5.b) [14].

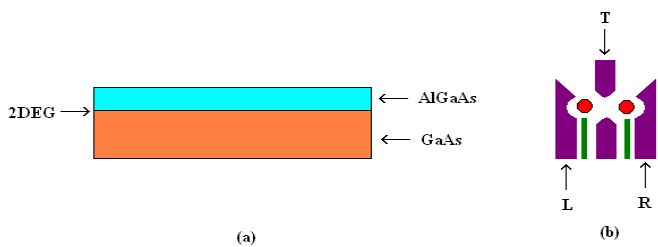

*Figure 1.5: (a) Schematic picture of a QD heterostructure. (b) Representation of a double QD. The red circles demonstrate the location of each QD and L, R and T are the metal gates.*

## 1.7.2 Energy structure of a double QD

As we know, electrons in real atoms will fill up the quantized energy levels according to the rules of quantum mechanics. The situation for electrons in QDs is the same as real atoms where the energy eigenstates in the potential wells are quantized. Therefore electrons will occupy the energy levels based on their coulomb interaction and their spin interaction. In order to study the lowest energy levels of the double QD we assume that there is only one electron in each QD and we proceed to study the relevant phenomena.

Denote $\varepsilon$ as the detuning, which is potential difference between the two QDs. In the absence of tunneling between the two QDs, electrons are highly localized in separated potential wells, and so we do not expect the two electrons to interact. This causes degeneracy in the energy of spin singlet state and spin triplet state. Figure 1.6(a) demonstrates this degeneracy where the numbers in the bracket indicate the number of electrons in each QD and S and T represent the spin singlet and triplet states respectively [14].

The spin singlet-triplet degeneracy is lifted by inter dot coupling due to electron tunneling through the potential barrier between the two QDs. The energy difference between these two states depends on the electrostatic charging energy of the double QD and the tunneling energy. This energy difference (J) has been illustrated in figure 1.6(b) [14].

From the energy-detuning diagram in figure 1.6(b), two suitable energy levels for encoding a qubit are 0 and 1. On one hand the insensitivity of the energy difference between 0 and 1 to the detuning variations at this spot makes this region suitable for qubit gate operation [12], but on the other hand the short spin relaxation time T1 (10-100 ns) in the absence of the external magnetic field [14] may undermine this scheme as we are aware that a long coherence time is a key requirement for building a scalable quantum computer.

The spin relaxation time for GaAs double QD has been extended to hundreds of milliseconds by applying a static magnetic field in plane of the double QD which makes it a very promising candidate for building scalable quantum computer. The energy degeneracy of the spin triplet state (1,1)T is lifted by applying an external magnetic field to the double QD, splitting the energy of three spin triplet states $(1,1)T_+$, $(1,1)T_0$, $(1,1)T_-$ to their Zeeman energies $E_Z$, where $T_+$ indicates that both electrons are in the spin up states, $T_-$ indicates that both electrons are in the spin down states and $T_0$ is the triplet state as a superposition of the spin up and spin down states. Spin states of this artificial atom are written as [14]

$$|(1,1)S\rangle = (|\uparrow\downarrow\rangle - |\downarrow\uparrow\rangle)/\sqrt{2} \qquad |(1,1)T_0\rangle = (|\uparrow\downarrow\rangle + |\downarrow\uparrow\rangle)/\sqrt{2} \qquad (1.24)$$

$$|(1,1)T_+\rangle = |\uparrow\uparrow\rangle \qquad |(1,1)T_-\rangle = |\downarrow\downarrow\rangle \qquad (1.25)$$

Here $|\uparrow\downarrow\rangle$ indicates spin up for the left QD and spin down for the right QD [14]. The environmental fluctuation that causes phase randomization of the electron spin in the double QD is another challenge for building a scalable quantum computer. One of the sources of spin decoherence in double QD is the hyperfine interaction that couples electronic and nuclear spins. The spin dephasing time due to this process has been extended beyond 1μs [15] and this suggests that GaAs QD can be a viable structure to be employed for constructing scalable quantum computer [14].

Another promising feature of the double QD is the capability of achieving the so called Quantum Non-Demolition (QND) measurement. The read-out process is performed by transferring one electron from one dot to another rather than losing it due to tunneling to the reservoir. We initially prepare the qubit in the $|(0,2)S\rangle$ state, then transfer one electron to another dot; after qubit manipulation by changing the voltage on the gates, the electron is returned to the initial QD but if during the qubit manipulation the spin of the electron is flipped, then it is forced to remain in the other QD due to Pauli's exclusion principle, otherwise the electron will be transferred to the first QD and the qubit will be in the $|(0,2)S\rangle$ state. In both cases the QND measurement is achieved, since we have not lost any of two electrons during the QC [14].

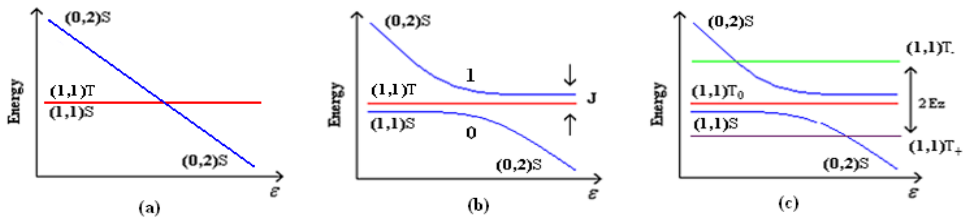

*Figure 1.6: (a) Energy diagram of a two electron double QD in the absence of the electron tunneling. (b) In the presence of the electron tunneling. (c) In the presence of the electron tunneling and applied magnetic field.*

# 2. The Geometric phase gate

In second quantization that we quantize energy levels of the free space, the number of modes (energy levels) where photons can occupy will reach to infinity. As we know in order to realize a fast qubit gate operation, the atoms have to be coupled strongly to the source of photons (strong coupling regime). In cavity QED we can achieve strong coupling by confining photons to a certain region of space (i.e. the cavity) with a limited number of modes. Consequently, the atoms will be coupled to a limited number of modes, increasing the atom-cavity coupling strength. In a thermal cavity, the intensity of the photons at different modes will vary as the temperature of the cavity changes.

In this chapter we will theoretically analyze a proposal to realize a C-Phase gate with a thermal cavity by appealing to a special evolution operator. Before going into the specific details of the scheme, we will introduce the basis of the model. Throughout this chapter, it will be assumed that qubits are embedded in atoms, and so qubit and atomic states are considered to be interchangeable.

## 2.1 Mathematical derivation of a geometric phase gate

The time evolution of an operator by applying the Hamiltonian $\hat{H}$ in the Heisenberg picture is given by [16]

$$\frac{d\hat{O}(t)}{dt} = i[\hat{H}, \hat{O}] \implies \hat{O}(t) = e^{i\hat{H}t}\,\hat{O}(0)\,e^{-i\hat{H}t} \tag{2.1}$$

Consider the commutation relation between dimensionless position and momentum operators $[\hat{X}, \hat{P}] = i$. If we select either the position or the momentum as the Hamiltonian, then the time evolution can be described as a displacement in the momentum or the position by an amount t, where t is the period of time that the Hamiltonian acts on the quantum system. Such evolution can be illustrated as a geometric path in the phase space.

Bearing in mind the characteristic behavior of the position and the momentum operators, we apply two Hamiltonians, $\hat{H}_1$ and $\hat{H}_2$ for the period of time t in the following order: $\hat{H}_1, \hat{H}_2, -\hat{H}_1, -\hat{H}_2$ so the evolution operator is given by [16]

$$\widehat{U}(t) = e^{i\widehat{H}_2 t} e^{i\widehat{H}_1 t} e^{-i\widehat{H}_2 t} e^{-i\widehat{H}_1 t} \qquad (2.2)$$

The following Taylor expansion is called the Campbell-Baker-Hausdorff (CBH) formula [17]

$$e^{i\widehat{H}_1 t} \widehat{H}_2 e^{-i\widehat{H}_1 t} = \widehat{H}_2 + it[\widehat{H}_1, \widehat{H}_2] - t^2/2! \, [\widehat{H}_1, [\widehat{H}_1, \widehat{H}_2]] + \ldots \qquad (2.3)$$

So according to the CBH formula we can rewrite (2.2) as

$$\widehat{U}(t) = e^{i\widehat{H}_2 t}\big(e^{i\widehat{H}_1 t} e^{-i\widehat{H}_2 t} e^{-i\widehat{H}_1 t}\big) = e^{i\widehat{H}_2 t}\big(e^{-i\widehat{H}_2 t} + it[\widehat{H}_1, e^{-i\widehat{H}_2 t}] - \cdots \big) =$$

$$1 + it\big(e^{i\widehat{H}_2 t} \widehat{H}_1 e^{-i\widehat{H}_2 t} - \widehat{H}_1\big) + \cdots = 1 + t^2[\widehat{H}_1, \widehat{H}_2] + \cdots \qquad (2.4)$$

Considering Taylor expansion we can write the evolution operator as [16]

$$\widehat{U}(t) = e^{[\widehat{H}_1, \widehat{H}_2] t^2} + O(t^3) \qquad (2.5)$$

where $O(t^3)$ denotes terms that are of order $t^3$ or higher.

This indicates that applying the Hamiltonian $i[\widehat{H}_1, \widehat{H}_2]$ has a similar effect to applying a set of Hamiltonians $\widehat{H}_1, \widehat{H}_2, -\widehat{H}_1, -\widehat{H}_2$, provided these Hamiltonians are only applied for a short time. The higher order terms in (2.5) consist of terms such as $[\widehat{H}_1, [\widehat{H}_1, \widehat{H}_2]]$, which vanish if $\widehat{H}_1, \widehat{H}_2$ are chosen so that they commute with $[\widehat{H}_1, \widehat{H}_2]$ and consequently CBH formula is an exact expression no matter how long the Hamiltonians are applied for.

Let's consider the following interaction Hamiltonian and investigate its evolution in phase space

$$\widehat{H} = \widehat{H}_1 + \widehat{H}_2 \qquad \widehat{H}_1 = a(t)\widehat{X}\widehat{S} \qquad \widehat{H}_2 = b(t)\widehat{P}\widehat{S} \qquad [\widehat{H}_1, \widehat{H}_2] = ia(t)b(t)\widehat{S}^2 \qquad (2.6)$$

$\widehat{S}$ is an operator that commutes with both $\widehat{X}$ and $\widehat{P}$, and $a(t)$ and $b(t)$ are arbitrary functions. Our aim is to apply this Hamiltonian for a certain period of time so that the evolution of the quantum state at the end of operation is independent of the time evolution of the position and the momentum. We can describe $\widehat{H}_1$ and $\widehat{H}_2$ as the position and the momentum dependent Hamiltonians where the time evolution of $\widehat{X}$ and $\widehat{P}$ in the Heisenberg picture will be [16]

$$\frac{d\widehat{X}(t)}{dt} = i[\widehat{H}, \widehat{X}] = b(t)\widehat{S} \implies \widehat{X}(T) - \widehat{X}(0) = \widehat{S} \int_0^T b(t)\, dt = \widehat{S} B(T) \qquad (2.7)$$

$$\frac{d\widehat{P}(t)}{dt} = i[\widehat{H}, \widehat{P}] = -a(t)\widehat{S} \implies \widehat{P}(T) - \widehat{P}(0) = -\widehat{S} \int_0^T a(t)\, dt = -\widehat{S} A(T) \qquad (2.8)$$

Now we choose the initial condition to make $\widehat{X}(0) = \widehat{P}(0) = 0$ which is the origin in phase space (figure 2.1), while the time is passing, the points $(B(t), -A(t))$ can trace a path through phase space.

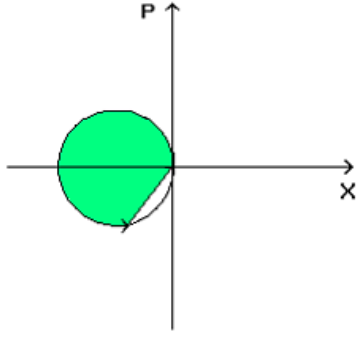

*Figure 2.1: Demonstration of the time evolution of a quantum state in the phase space. The colored area is determined by C(t) after applying the Hamiltonian for the period of time t.*

The evolution operator for these Hamiltonians is [9]

$$\hat{U}(t) = e^{-iC(t)\hat{S}^2} e^{-iA(t)\hat{X}\hat{S}} e^{-iB(t)\hat{P}\hat{S}} \quad (2.9)$$

We can obtain $C(t)$ by applying the Schrödinger equation [9]

$$\frac{d\hat{U}(t)}{dt} = -i\hat{H}\hat{U}(t) \quad [\hat{S}, \hat{X}] = [\hat{S}, \hat{P}] = 0 \quad => \quad (2.10)$$

$$\left(\frac{dC(t)}{dt}\hat{S}^2 + \frac{dA(t)}{dt}\hat{X}\hat{S}\right)\hat{U}(t) + e^{-iC(t)\hat{S}^2} e^{-iA(t)\hat{X}\hat{S}} \left(\frac{dB(t)}{dt}\hat{P}\hat{S}\right) e^{-iB(t)\hat{P}\hat{S}} = \left(a(t)\hat{X}\hat{S} + b(t)\hat{P}\hat{S}\right)\hat{U}(t) \quad (2.11)$$

The CBH formula implies that

$$e^{-iA(t)\hat{X}\hat{S}} \left(\frac{dB(t)}{dt}\hat{P}\hat{S}\right) e^{iA(t)\hat{X}\hat{S}} e^{-iA(t)\hat{X}\hat{S}} = \frac{dB(t)}{dt}\left(\hat{P} + A(t)\hat{S}\right)\hat{S} e^{-iA(t)\hat{X}\hat{S}} \quad (2.12)$$

substituting (2.12) into (2.11) gives

$$\left(\frac{dC(t)}{dt}\hat{S}^2 + \frac{dB(t)}{dt}A(t)\hat{S}^2 + \frac{dA(t)}{dt}\hat{X}\hat{S} + \frac{dB(t)}{dt}\hat{P}\hat{S}\right)\hat{U}(t) = \left(a(t)\hat{X}\hat{S} + b(t)\hat{P}\hat{S}\right)\hat{U}(t) \quad (2.13)$$

which, together with (2.7) and (2.8) gives

$$C(T) = -\int_0^T A(t)b(t)\,dt \quad (2.14)$$

From (2.14), we can see that the traversed path in phase space will enclose an area as depicted in figure 2.1 [9]. This area represents the amount of phase acquired by the quantum state during the time evolution (T), which is determined by $C(t)$. We can choose the suitable functions to satisfy the condition $(B(T) = A(T) = 0)$ and, by applying the Hamiltonian for the right amount of time (T), we can traverse a closed path in phase space and return to the origin, where the only limitation on the operation time is that it must be shorter than the coherence time.

The geometric phase gate has been successfully demonstrated in a two-qubit ion trap experiment where a geometric phase gate has been achieved independent of the motional state of the ions [10]. The thermal vibration of the ions is the result of absorption and emission of photons via interactions with the cavity field, which can be represented by the photon creation and annihilation operators in the interaction Hamiltonian. These terms in the interaction Hamiltonian can be replaced by the position and the momentum operator to describe the motional state of ions.

In order to manipulate the qubit states (internal quantum states of the two ions) for quantum gate operation, ions are illuminated with two different classical fields (laser fields). These fields exert an oscillating force on the ions. The frequency detuning of the laser beams will cause a time-dependent phase difference between the two qubit states. By adjusting the frequency detuning of the laser beams and applying them for a certain amount of time, the qubit states will return to their initial states, independent of the thermal vibrations [10].

In the following section we will theoretically explain how we can exploit this successful technique to perform a C-Phase gate in cavity QED where the atoms will be fluctuating due to interaction with the thermal field. A classical field is applied to the atoms and by satisfying some conditions on the Rabi frequency and time evolution of the system, we can realize a C-Phase gate.

## 2.2 Realization of a two-qubit C-Phase gate with a thermal cavity

In this section we will focus on the physical principles behind the realization of the C-Phase gate. Interested readers can find comprehensive derivations of the equations in appendix B.

Atoms in a thermal cavity which are in a dispersive coupling regime can interact with the cavity by exchanging virtual photons, randomizing the phase of the atomic states. Realizing a C-Phase gate will be prevented by thermal fluctuations, unless, as before, the interaction Hamiltonian is replaced by another Hamiltonian which has a time evolution that is unaffected by thermal fluctuations.

We will overcome the problem by applying a strong classical field [18] (large amplitude oscillating electromagnetic field) in resonance with the atomic transition frequency so that the atomic state will oscillate between the ground and excited states with the Rabi frequency, which will increase with the intensity of the applied field. It is essential to apply the classical field to both of the atoms at the same time. The atomic transition frequency is shifted because of the oscillating electric field, this is referred to as the *ac-Stark shift* [19]; this causes a dephasing on its own. We will be able to eliminate the terms responsible for this dephasing by choosing a high intensity field. Because of the frequency difference between the atom-cavity detuning and the Rabi frequency, atomic states will accumulate a phase over time. This phase accumulation can be visualized in the phase space diagram since we can relate the photon creation and annihilation operators to the dimensionless position and momentum operators by [9]

$$\hat{a}^\dagger = 1/\sqrt{2}(\hat{X} - i\hat{P}) \qquad \hat{a} = 1/\sqrt{2}(\hat{X} + i\hat{P}) \qquad (2.15)$$

The phase accumulation is a time dependent phenomenon, so by choosing appropriate initial conditions and considering the coherence time of the cavity we will choose the operation time in such a way that the atomic state will return to its initial state in phase space, with an overall phase factor that depends on the initial atomic state. This means we can realize a C-Phase gate in a thermal cavity where the atomic states will acquire the appropriate amount of phase because of the phase accumulation during the operation time.

The free Hamiltonian describing a single mode cavity and two identical two level atoms can be represented by (with $\hbar = 1$) [18]

$$\hat{H}_{free} = \hat{H}_{atom} + \hat{H}_{cavity} = \omega_0 \hat{S}_z + \omega_a \hat{a}^\dagger \hat{a} \qquad (2.16)$$

$$\hat{S}_z = 1/2 \, \Sigma_{j=1,2}(|e_j\rangle\langle e_j| - |g_j\rangle\langle g_j|) \qquad (2.17)$$

$\omega_0$ is the transition frequency of the atom, $\omega_a$ is the frequency of the cavity mode and atoms have been labeled by the index j.

By using the JC Hamiltonian and the rotating wave approximation, we can describe the simultaneous interaction between two atoms and a single mode cavity in the presence of a class classical field as [18]

$$\hat{H}_i = \hat{H}_Q + \hat{H}_C = g/2\Sigma_{j=1,2}(\hat{a}^\dagger \hat{S}_j^- + \hat{a}\hat{S}_j^+) + \Omega/2\Sigma_{j=1,2}\left(\hat{S}_j^+ e^{-i\omega t} + \hat{S}_j^- e^{i\omega t}\right) \qquad (2.18)$$

$$\hat{S}_j^+ = |e_j\rangle\langle g_j| \qquad \hat{S}_j^- = |g_j\rangle\langle e_j| \qquad (2.19)$$

Where $\hat{H}_i$ is the interaction Hamiltonian, $\hat{H}_Q$ describes the interaction of the two atoms with the quantized cavity mode, and g is the atom-cavity coupling constant. The interaction between the atoms and a classical field is expressed by $\hat{H}_C$, where $\Omega$ is the Rabi frequency and $\omega$ is frequency of the classical field. Hence the total Hamiltonian for such structure is [18]

$$\hat{H}_{total} = \hat{H}_i + \hat{H}_{free} \qquad (2.20)$$

Considering the two qubits in resonance with the classical field ($\omega_0 = \omega$) and denoting the atom-cavity detuning ($\delta = \omega_0 - \omega_a$), we use the free Hamiltonian to transform into the interaction picture. In the interaction picture, the interaction Hamiltonian is [18]

$$\hat{H}_{int} = \Sigma_{j=1,2}[g/2(\hat{a}^\dagger \hat{S}_j^- e^{-i\delta t} + \hat{a}\hat{S}_j^+ e^{i\delta t}) + \Omega/2(\hat{S}_j^+ + \hat{S}_j^-)] \qquad (2.21)$$

The first part of this Hamiltonian describes the atoms dispersive coupling to the cavity and the second part indicates that the atomic states which are in resonance with the classical field are oscillating with the Rabi frequency. Since the atoms are not in resonance with the cavity, the new basis states can be defined as the superposition of the ground state and the excited state [18]

$$|+_j\rangle = 1/\sqrt{2}(|g_j\rangle + |e_j\rangle) \qquad |-_j\rangle = 1/\sqrt{2}(|g_j\rangle - |e_j\rangle) \qquad (2.22)$$

In this new basis, the interaction Hamiltonian can be written as [18]

$$\hat{H}_{int} = \Sigma_{j=1,2}\{g/2[\hat{a}^\dagger e^{-i\delta t}(\hat{\sigma}_j^z + 1/2\hat{\sigma}_j^- - 1/2\hat{\sigma}_j^+) +$$

$$\hat{a}e^{i\delta t}(\hat{\sigma}_j^z + 1/2\hat{\sigma}_j^+ - 1/2\hat{\sigma}_j^-)] + \Omega\hat{\sigma}_j^z\} \qquad (2.23)$$

In order to remove the time-independent term from the Hamiltonian we can write the interaction Hamiltonian in a frame that rotates with the angular frequency $\Omega$. First we find the time evolution of the system in the interaction picture by applying the Schrödinger equation [18]

$$i\frac{d|\psi(t)\rangle}{dt} = \widehat{H}_{int}|\psi(t)\rangle \qquad (2.24)$$

Next we investigate the time evolution of the system in the Schrödinger picture where by inspecting (2.23) it is easy to see that there is a time-independent part in the interaction Hamiltonian $\left(\widehat{H}_0 = \Omega\Sigma_{j=1,2}\widehat{\sigma}_j^z\right)$. Then it can be used to find the time evolution of the system ($|\psi'(t)\rangle$) in the Schrödinger picture [18]

$$|\psi(t)\rangle = e^{-i\widehat{H}_0 t}|\psi'(t)\rangle \qquad (2.25)$$

From (2.23)-(2.25) we find the interaction Hamiltonian in the Schrödinger picture to be

$$\widehat{H}_i' = g/2\Sigma_{j=1,2}[\hat{a}^\dagger e^{-i\delta t}(\widehat{\sigma}_j^z + 1/2\widehat{\sigma}_j^- e^{-i\Omega t} - 1/2\widehat{\sigma}_j^+ e^{i\Omega t}) +$$
$$\hat{a}e^{i\delta t}(\widehat{\sigma}_j^z + 1/2\widehat{\sigma}_j^+ e^{i\Omega t} - 1/2\widehat{\sigma}_j^- e^{-i\Omega t})] \qquad (2.26)$$

We have separated the time-independent part of the interaction Hamiltonian for clarity. In order to eliminate the ac-Stark shift terms we apply the classical field in such a way that the Rabi frequency satisfies $\Omega \gg \delta, g$, then we will be able to simplify the interaction Hamiltonian by ignoring the fast oscillating terms in the RWA to obtain [18]

$$\widehat{H}_i' = g/2\Sigma_{j=1,2}(\hat{a}^\dagger e^{-i\delta t} + \hat{a}e^{i\delta t})\widehat{\sigma}_j^z = g/2(\hat{a}^\dagger e^{-i\delta t} + \hat{a}e^{i\delta t})\widehat{S}_x \qquad (2.27)$$

The similarity between (2.6) and (2.27) and the fact that $[\hat{a}, \hat{a}^\dagger] = \mathbf{I}$ suggests that we will be able to find an evolution operator that is independent of thermal fluctuations. Therefore we can write the evolution operator for the Hamiltonian (2.27) as [18]

$$\widehat{U}'(t) = e^{-iA(t)\widehat{S}_x^2} e^{-iB(t)\widehat{S}_x\hat{a}} e^{-iC(t)\widehat{S}_x\hat{a}^\dagger} \qquad \widehat{H}_i' = (\hat{a}^\dagger c(t) + \hat{a}b(t))\widehat{S}_x \qquad (2.28)$$

In order to find the time dependent functions we apply the Schrödinger equation [18]

$$\frac{d\widehat{U}'(t)}{dt} = -i\widehat{H}_i'\widehat{U}'(t) \qquad (2.29)$$

We choose the initial conditions to be A(0)=B(0)=C(0)=0, since the atoms and the cavity will become entangled during the operation time and so the cavity must not decay during the operation, otherwise the entanglement would be lost. If we know the atom-cavity detuning, then by choosing the minimum operation time to satisfy $\delta t = 2\pi$ we can write the evolution operator in the interaction picture as [18]

$$\widehat{U}(t) = e^{-i\widehat{H}_0 t}\widehat{U}'(t) = e^{-i\Omega t\widehat{S}_x - i\lambda t\widehat{S}_x^2} \qquad \lambda = g^2/4\delta \qquad (2.30)$$

It is evident that the time evolution of the system is independent of the photon creation and annihilation operators, which means that the atoms can evolve independent of the thermal fluctuation as long as we satisfy the aforementioned conditions.

Now we apply the evolution operator to atomic states to find out whether we can realize a C-Phase gate or not. Therefore we consider the following relations

$$\hat{S}_x|+_1\rangle|+_2\rangle = |+_1\rangle|+_2\rangle, \hat{S}_x|+_1\rangle|-_2\rangle = \hat{S}_x|-_1\rangle|+_2\rangle = 0, \hat{S}_x|-_1\rangle|-_2\rangle = -|-_1\rangle|-_2\rangle \quad (2.31)$$

$$\hat{S}_x^2|+_1\rangle|+_2\rangle = |+_1\rangle|+_2\rangle, \hat{S}_x^2|+_1\rangle|-_2\rangle = \hat{S}_x^2|-_1\rangle|+_2\rangle = 0, \hat{S}_x^2|-_1\rangle|-_2\rangle = |-_1\rangle|-_2\rangle \quad (2.32)$$

From (2.30), (2.31) and (2.32) we can write

$$\hat{U}(t)|+_1\rangle|+_2\rangle = e^{-i(\Omega+\lambda)t}|+_1\rangle|+_2\rangle \quad (2.33)$$

$$\hat{U}(t)|+_1\rangle|-_2\rangle = |+_1\rangle|-_2\rangle \quad (2.34)$$

$$\hat{U}(t)|-_1\rangle|+_2\rangle = |-_1\rangle|+_2\rangle \quad (2.35)$$

$$\hat{U}(t)|-_1\rangle|-_2\rangle = e^{i(\Omega-\lambda)t}|-_1\rangle|-_2\rangle \quad (2.36)$$

According to these relations, to realize a C-Phase gate we also need to satisfy [18]

$$g = \delta \Rightarrow \lambda t = tg^2/4\delta = \pi/2 \qquad \Omega t = (2n + 1/2)\pi \Rightarrow$$

$$(\Omega - \lambda)t = 2n\pi \qquad (\Omega + \lambda)t = (2n + 1)\pi \quad (2.37)$$

where n is an integer, and bearing in mind that we need to have $\Omega \gg \delta, g$. Based on these conditions, we have [18]

$$\hat{U}(t)|+_1\rangle|+_2\rangle = -|+_1\rangle|+_2\rangle \quad (2.38)$$

$$\hat{U}(t)|+_1\rangle|-_2\rangle = |+_1\rangle|-_2\rangle \quad (2.39)$$

$$\hat{U}(t)|-_1\rangle|+_2\rangle = |-_1\rangle|+_2\rangle \quad (2.40)$$

$$\hat{U}(t)|-_1\rangle|-_2\rangle = |-_1\rangle|-_2\rangle \quad (2.41)$$

That is, $\hat{U}(t)$ is a C-Phase gate.

## 2.3 Summary

In summary we have demonstrated that the time evolution of a quantum system through a sequence of the Hamiltonians can be achieved similarly by replacing those Hamiltonians with another Hamiltonian under certain conditions. The new Hamiltonian will allow the time evolution of a system to be controlled so that a geometric phase gate can be implemented. The geometric phase gate is achieved by choosing the appropriate time dependent functions in the Hamiltonian and operating the Hamiltonian for the correct amount of time. The atomic state traverses a path in the phase space and will return to its initial state, having accumulated a phase depending on the initial atomic state. The overall evolution of the geometric phase gate is equivalent to a C-Phase gate.

This technique has been demonstrated successfully in an ion trap experiment to achieve a geometric phase gate independent of the motional state of the ions. Later on this method was extended to cavity QED, where by controlling the intensity of the applied classical field we could ignore the ac-Stark shift effect from the Hamiltonian. As a result of controlling the time evolution of the system and the Rabi frequency we realized a c-Phase gate with thermal cavity.

The implementation of the C-Phase gate with a thermal cavity in cavity QED indicates that we can generalize this method to circuit QED. In a circuit QED experiment, we would place the double QDs (qubits) in the vicinity of a TLR, where we will realize the qubit entanglement via capacitive coupling of the qubits to the TLR. By adjusting the voltage difference between the gates we could control the transition frequency of the double QDs. Therefore it will be possible to perform the gate operation reliably, even in the presence of thermal fluctuations.

In circuit QED we will replace the thermal cavity with a TLR which will act as the cavity mediated coupling between the double QDs to realize the entanglement. We will apply an oscillating voltage to one of the gates on the double QDs and by controlling the Rabi frequency of the applied field as well as operation time we will realize a C-Phase gate.

# 3. Generation of the cluster states in circuit QED

Quantum Information Processing coupling double QDs to a TLR was introduced by Childress *et al*. [20]. In their scheme, an electron would be shared between two neighboring QDs, where the spin-charge states are considered as the energy eigenstates of the double QD. Since the charge states have short coherence times (~ns) these states were not suitable for QIP; hence they proposed to entangle the spin states with the charge states because electrons have a long spin relaxation time in the presence of an applied magnetic field [14].

Although there were other proposals to perform QIP by spin qubit manipulation [5][21] the idea of capacitive coupling between the double QDs and a TLR was further developed by Taylor & Lukin [22] who employed two electrons in a double QD to use the spin degree of freedom.

## 3.1 An overview of the proposal

In this chapter we demonstrate theoretically that a cluster state can be generated by performing a C-Phase gate between double QDs based on the same technique that we discussed in chapter two. It is important to note that part of our technique is the generalization of the proposal made by Taylor & Lukin [22].

In this scheme we consider the artificial atoms (double QDs) as the qubits, which replace the atoms used in cavity QED. We employ the TLR as the cavity mediated coupling between the double QDs, qubits are encoded in the charge states for the qubit manipulation purpose.

We consider two electrons in the double QDs where the lowest energy level for the doubly occupied QD is the spin singlet state $|(0,2)S\rangle$. The double QDs are placed in the vicinity of the TLR in such a way that one of the dots is coupled to the TLR via capacitive coupling [23].

First we initialize the qubits in the ground state. Then by applying a pulse to the gates, we will prepare them in the ground state of the new charge basis states (a superposition of the charge basis states). The capacitive coupling of qubits to the TLR in dispersive mode will cause the phase randomization of the qubit states. This is similar to the situation that we had in the thermal cavity where the atom-cavity interaction would take place via virtual photon exchange. Therefore we will be able to achieve the C-Phase gate by using the same method that we employed in cavity QED [23].

An oscillating voltage is applied to one of the gates in resonance with transition frequency of the charge qubits; this process is analogous to applying the classical field in cavity QED. The transition frequency of the qubits will be adjusted to be almost resonant with the fundamental mode of the TLR so the qubit-TLR coupling will be maximized. We will apply the oscillating voltage for the right amount of time then by imposing a condition on the Rabi frequency and the qubit-TLR frequency detuning we will realize a C-Phase gate [23].

As discussed in section 1.2.1, the cluster state associated to the 1D array of qubits is created by preparing the qubits in the state $|-_i\rangle$ and performing the C-Phase gate between them. Since we satisfied some conditions to achieve the C-Phase gate, this 1D array of qubits can be transformed into a cluster state.

Coherence time of the charge qubits at the avoided crossing will be short, so in order to accomplish the single qubit readout process we will transfer qubits to the large positive detuning region where $|-\rangle$ & $|+\rangle$ states are mapped to $|(0,2)S\rangle$ & $|(1,1)S\rangle$ states respectively. A large static magnetic field will be applied in the plane of the 2DEG to increase the spin relaxation time [14] hence we can prevent the spin singlet-triplet mixing in that region [23].

## 3.2 Hamiltonian for a Double QD

A double QD is formed by confining electrons to a certain area of the 2DEG, where they are localized in the potential wells. This process is achieved by applying a voltage to gates on top of the heterostructure so that an electric field is set up under the gates. Due to the large dipole moment of the double QD, we can expect strong coupling between the double QD and the gates, allowing quantum gates to be applied on short timescales.

We intend to exploit the lowest energy level of the two-electron double QDs for QIP. Therefore we choose $|(0,2)S\rangle$ and $|(1,1)S\rangle$ as the ground and excited states (charge basis states) respectively; excitation from the ground state to the excited state takes place by transferring an electron from one dot to another via a tunneling mechanism (figure 3.1). Tunneling can be controlled by adjusting the bias voltage on the gates. The excitation time is inversely proportional to the tunneling rate T.

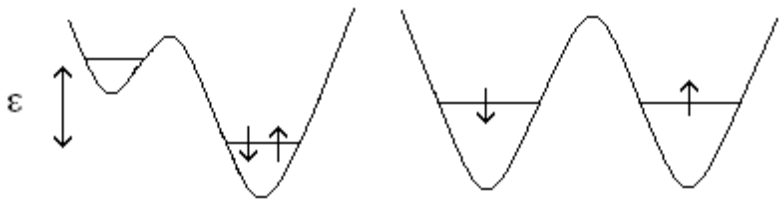

*Figure 3.1: Illustration of the electron tunneling between two QDs, ε is the potential energy difference between two QDs. Arrows demonstrate the spin orientation of the electrons.*

The degeneracy of the singlet-triplet state is lifted due to tunneling between the QD. The Hamiltonian for this three level system can be written as [12]

$$\hat{H} = E_0|(1,1)S\rangle\langle(1,1)S| + E_1|(1,1)T_0\rangle\langle(1,1)T_0| +$$

$$T(|(1,1)S\rangle\langle(0,2)S| + |(0,2)S\rangle\langle(1,1)S|) + \varepsilon|(0,2)S\rangle\langle(0,2)S| \tag{3.1}$$

where $\varepsilon$ is the potential energy detuning between two dots and 2T is the tunneling energy. In order to manipulate the charge states we will transfer one electron from a doubly occupied QD to the other QD by applying a pulse to one of the gates.

The energy required to transfer one electron from one potential well to another will be the electrostatic charging energy. Considering the large energy gap of the singlet-triplet state in the doubly occupied QD, we can regard the Coulomb interaction as the dominant interaction and so we can ignore the spin interaction energies ($E_0 \approx E_1 = 0$) in the Hamiltonian [24].

The eigenstates for the Hamiltonian in the rotated frame of the charge basis states can be written as [23]

$$|+\rangle = -sin\theta|(1,1)S\rangle + cos\theta|(0,2)S\rangle \qquad |-\rangle = cos\theta|(1,1)S\rangle + sin\theta|(0,2)S\rangle \tag{3.2}$$

The new basis states are hybridized charge states where $\theta$ is referred to as the adiabatic angle [24]. The relation between the adiabatic angle and the energy parameters is determined by obtaining the eigenvalues of the Hamiltonian [24]

$$\lambda_1 = \frac{\varepsilon - \sqrt{\varepsilon^2 + 4T^2}}{2} \qquad \lambda_2 = \frac{\varepsilon + \sqrt{\varepsilon^2 + 4T^2}}{2} \qquad tan\theta = \frac{-2T}{\varepsilon + \sqrt{\varepsilon^2 + 4T^2}} \tag{3.3}$$

where $\omega_d = \sqrt{\varepsilon^2 + 4T^2}$ is transition frequency of the new charge states. The qubit entanglement is achieved by coupling the double QDs to the TLR, where the coupling strength will play a crucial role. The adiabatic angle is an arbitrary parameter, but, we set $\theta = -\pi/4$ for two reasons.

This value represents the operation area where the qubit-TLR coupling is maximized and the transition frequency at this spot will be insensitive to first order variations in the detuning [22]. This choice represents the avoided crossing region in energy diagram of the double QD (figure 3.2). To simplify notation, we set the detuning $\varepsilon = 0$ in this point [24].

Hence we can define the new charge basis states at the avoided crossing as [23]

$$|+\rangle = 1/\sqrt{2}(|(1,1)S\rangle + |(0,2)S\rangle) \qquad |-\rangle = 1/\sqrt{2}(|(1,1)S\rangle - |(0,2)S\rangle) \tag{3.4}$$

The interaction Hamiltonian for this system consists of two parts. Firstly we are required to write the interaction Hamiltonian for interaction of the double QDs with the gates implemented above the heterostructure. Then we need to describe the interaction of the double QDs with the TLR. The total interaction Hamiltonian will be the sum of these two Hamiltonians.

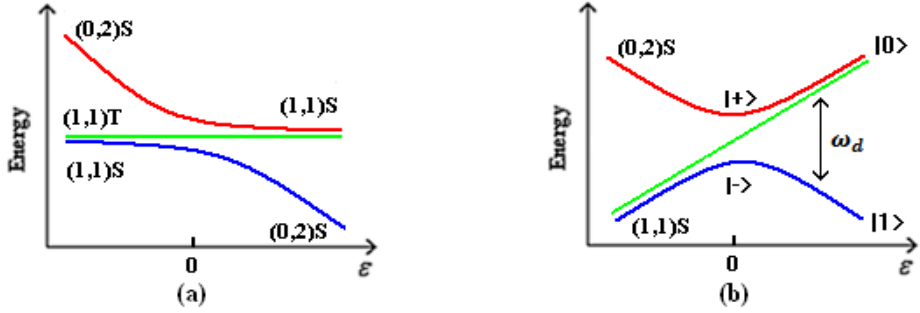

*Figure 3.2: (a) The lowest energy levels of the two-electron double QD. (b) Energy levels in the rotated frame of the charge basis states.*

The interaction between the electric dipole moment of the double QD and the electric field produced by the gate can be described by the following Hamiltonian [23]

$$\hat{H}_{dot} = \eta |(1,1)S\rangle\langle(1,1)S| \qquad \eta = -\vec{d}.\vec{E} \qquad (3.5)$$

where $\eta$ is the electric dipole moment interaction energy, $\vec{d}$ is the electric dipole moment of the double QD and $\vec{E}$ is the electric field under the gate. The qubits are encoded in these charge states during the qubit manipulation process so we can define new operators and rewrite the Hamiltonian (3.5) for the *i*th qubit as

$$\hat{S}_i^+ = |+_i\rangle\langle-_i| \qquad \hat{S}_i^- = |-_i\rangle\langle+_i| \qquad \hat{S}_i^x = \hat{S}_i^+ + \hat{S}_i^- \qquad (3.6)$$

$$\hat{H}_{dot} = \eta/2(I + \hat{S}_i^x) = (\eta/2)\hat{S}_i^x \qquad (3.7)$$

We have ignored the constant energy term from the Hamiltonian as it will not conserve energy during the interaction with the TLR.

## 3.3 Realization of an Ising-like operator in circuit QED

We can achieve the C-Phase gate by coupling the qubits to a TLR resulting in entanglement between two qubits where the qubits are entangled. The Coulomb interaction between the qubits is minimized by placing them at distances larger than their size [23]. The low mode volume of 1D TLR allows us to treat this structure as a single mode cavity that allows strong coupling.

We place the double QDs adjacent to the TLR (figure 3.3), where the left QD is coupled to the TLR by a coupling capacitance $C_C$, $C_t$ is the total capacitance of the double QD including the coupling capacitance to the TLR. The voltages applied to the left QD and the TLR are

$$\hat{V}_{dot} = \frac{e}{2C_t}(I + \hat{S}_i^x) \qquad \hat{V}_{TLR} = \sqrt{\hbar\omega/Lc_0}(\hat{a} + \hat{a}^\dagger) \qquad (3.8)$$

respectively, where $\omega$ is the frequency of the fundamental mode of the TLR, $L$ and $c_0$ are the length and capacitance per unit length of the TLR respectively.

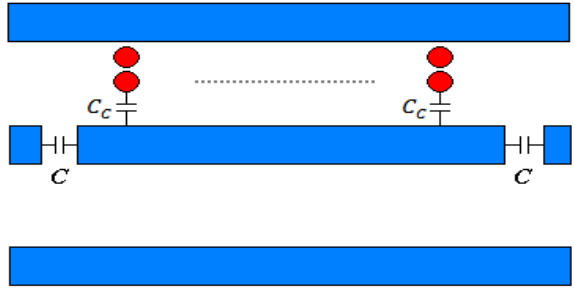

*Figure 3.3: Demonstration of the double QDs (red) capacitive coupling to a TLR (blue).*

The electrostatic charging energy of the double QDs will change by transferring one electron from one dot to another and the variation of charging energy will affect the interaction energy between the double QD and the TLR. As double QDs are capacitively coupled to the TLR, we can write the interaction Hamiltonian in the RWA as (with $\hbar = 1$) [12]

$$\hat{H}_i = C_C \hat{V}_{TLR} \hat{V}_{dot} = g_0 \Sigma_i (\hat{a}^\dagger \hat{S}_i^- + \hat{a} \hat{S}_i^+) \tag{3.9}$$

This is just the standard JC Hamiltonian that represents the simultaneous resonant interaction of several qubits with a quantized single mode cavity (TLR), $g_0$ is the coupling coefficient at the avoided crossing which is assumed to be similar for all the qubits [23]. Considering the electrostatic energy, $E_c$ required for transferring one electron from one dot to another we can describe the coupling coefficient by using the quantized resistance, $R_Q$, and the characteristic impedance of the TLR, $z_0$, [20]

$$g_0 = \omega \frac{C_C}{2C_t} \sqrt{\frac{2z_0}{R_Q}} \quad R_Q = \frac{h}{e^2} \approx 26 k\Omega \quad z_0 = 50\Omega \quad \omega = \frac{\pi}{Lc_0 z_0} \quad E_c = \frac{e^2}{C_t} \tag{3.10}$$

In order to find the coupling coefficient at any adiabatic angle we replace the basis states from (3.2) in (3.6) and rewrite (3.9) as

$$\hat{H}_i = C_C \hat{V}_{TLR} \hat{V}_{dot} = -g_0 \sin 2\theta \Sigma_i (\hat{a}^\dagger \hat{S}_i^- + \hat{a} \hat{S}_i^+) \tag{3.11}$$

Now we can use (3.3) to find the effective coupling coefficient as

$$g_{eff} = -g_0 \sin 2\theta \quad \sin 2\theta = \frac{2\tan\theta}{1+\tan^2\theta} = -\frac{2T}{\omega_d} \quad => \quad g_{eff} = g_0 \frac{2T}{\omega_d} \tag{3.12}$$

This indicates that coupling coefficient is maximized at the avoided crossing where transition frequency is minimized. This justifies our choice to operate at this spot in order to have stronger coupling to the TLR. The TLR will act as a cavity and according to figure 3.2(b) we can write the free Hamiltonian for this system as

$$\hat{H}_{free} = \omega \hat{a}^\dagger \hat{a} + \omega_d \Sigma_i \hat{S}_i^z \qquad \hat{S}_i^z = 1/2(|+_i\rangle\langle+_i| - |-_i\rangle\langle-_i|) \tag{3.13}$$

The free Hamiltonian is used to transform (3.9) to the interaction picture. In the interaction picture, the interaction Hamiltonian will be [23]

$$\hat{H}_{int} = g_0 \Sigma_i \left( e^{i\delta t} \hat{a}^\dagger \hat{S}_i^- + e^{-i\delta t} \hat{a} \hat{S}_i^+ \right) \qquad \delta = \omega - \omega_d \qquad (3.14)$$

This Hamiltonian describes the dispersive coupling between double QDs and the cavity where the phase randomization of the qubit states must be overcome to realize a C-Phase gate.

We will apply a high intensity oscillating electric field to the left gate of the double QDs simultaneously. The oscillating field will be in resonance with the qubit transition frequency at the operation point. So the electric dipole moment of the double QDs will oscillate with Rabi frequency $\eta$ and the interaction Hamiltonian including capacitive coupling to the cavity is

$$\hat{H}_{int} = \Sigma_i [g_0 \left( e^{i\delta t} \hat{a}^\dagger \hat{S}_i^- + e^{-i\delta t} \hat{a} \hat{S}_i^+ \right) + (\eta/2) \hat{S}_i^x] \qquad (3.15)$$

After applying the oscillating voltage for the right amount of time we will apply a pulse to the gates to the left QDs, mapping charge states to the spin singlet states in the far positive detuning region (figure 3.2(b)).

We define the new operators and the new basis states $|0\rangle$ and $|1\rangle$ for QIP

$$|0\rangle = |(1,1)S\rangle \qquad |1\rangle = -|(0,2)S\rangle \qquad \hat{\sigma}_i^z = |0_i\rangle\langle 0_i| - |1_i\rangle\langle 1_i| = \hat{S}_i^x \qquad (3.16)$$

$$\hat{\sigma}_i^- = |0_i\rangle\langle 1_i| \qquad \hat{\sigma}_i^+ = |1_i\rangle\langle 0_i| \qquad =>$$

$$\hat{S}_i^- = (1/2)(\hat{\sigma}_i^z - \hat{\sigma}_i^- + \hat{\sigma}_i^+) \qquad \hat{S}_i^+ = (1/2)(\hat{\sigma}_i^z + \hat{\sigma}_i^- - \hat{\sigma}_i^+) \qquad (3.17)$$

Substituting the new operators in (3.15) the interaction Hamiltonian will be

$$\hat{H}_{int} = \Sigma_i \{(g_0/2)[e^{i\delta t} \hat{a}^\dagger (\hat{\sigma}_i^z - \hat{\sigma}_i^- + \hat{\sigma}_i^+) + e^{-i\delta t} \hat{a}(\hat{\sigma}_i^z + \hat{\sigma}_i^- - \hat{\sigma}_i^+)] + (\eta/2)\hat{\sigma}_i^z\} \qquad (3.18)$$

In order to remove the time-independent term from the Hamiltonian we can write the interaction Hamiltonian in a frame that rotates with the angular frequency $\eta/2$. First we find the time evolution of the system in the interaction picture by applying the Schrödinger equation [18]

$$i\frac{d|\psi(t)\rangle}{dt} = \hat{H}_{int} |\psi(t)\rangle \qquad (3.19)$$

Next we investigate the time evolution of the system in the Schrödinger picture where by inspecting (3.18) it is easy to see that there is a time-independent part $(\hat{H}_0 = (\eta/2)\Sigma_i \hat{\sigma}_i^z)$ in the interaction Hamiltonian. This can be used to find the time evolution of the system ($|\psi'(t)\rangle$) in the Schrödinger picture [18]

$$|\psi(t)\rangle = e^{-i\hat{H}_0 t} |\psi'(t)\rangle \qquad (3.20)$$

From (3.18)-(3.20) we find the interaction Hamiltonian in the Schrödinger picture

$$\hat{H}_i' = (g_0/2)\Sigma_i [\hat{a}^\dagger e^{i\delta t}(\hat{\sigma}_i^z - \hat{\sigma}_i^- e^{i\eta t} + \hat{\sigma}_i^+ e^{-i\eta t}) + \hat{a} e^{-i\delta t}(\hat{\sigma}_i^z - \hat{\sigma}_i^+ e^{-i\eta t} + \hat{\sigma}_i^- e^{i\eta t})] \qquad (3.21)$$

We have simply separated the time-independent part of the interaction Hamiltonian for clarity. To eliminate the ac-Stark shift terms we apply the oscillating field in such a way that the Rabi frequency satisfies $\eta \gg \delta$, then we can simplify the interaction Hamiltonian by ignoring the fast oscillating terms

$$\hat{H}_i' = (g_0/2)\Sigma_i(\hat{a}^\dagger e^{i\delta t} + \hat{a}e^{-i\delta t})\hat{\sigma}_i^z \tag{3.22}$$

In order to realize a C-Phase gate we can use the same technique that we employed with the thermal cavity. We write the evolution operator for the interaction Hamiltonian as

$$\hat{U}'(t) = e^{-iA(t)(\Sigma_i\hat{\sigma}_i^z)^2}e^{-iB(t)\Sigma_i\hat{\sigma}_i^z\hat{a}}e^{-iC(t)\Sigma_i\hat{\sigma}_i^z\hat{a}^\dagger} \qquad \hat{H}_i' = (\hat{a}^\dagger c(t) + \hat{a}b(t))\Sigma_i\hat{\sigma}_i^z \tag{3.23}$$

We apply Schrödinger's equation to find the time dependent functions in the evolution operator; therefore we have [18]

$$\frac{d\hat{U}'(t)}{dt} = -i\hat{H}_i'\hat{U}'(t) \tag{3.24}$$

we choose the initial conditions to be A(0)=B(0)=C(0)=0. If the operation time satisfies $\delta t = 2k\pi$ (where $k$ is integer) then the evolution operator in the interaction picture will be [23]

$$\hat{U}'(t) = exp[-i\lambda/2t(\Sigma_i\hat{\sigma}_i^z)^2] \qquad \lambda = g_0^2/2\delta \tag{3.25}$$

As $|0_i\rangle$ and $|1_i\rangle$ are eigenstates of the $\hat{\sigma}_i^z$ operators, we can write

$$\hat{\sigma}_i^z|0_i\rangle = |0_i\rangle \qquad \hat{\sigma}_i^z|1_i\rangle = -|1_i\rangle \qquad (\hat{\sigma}_i^z)^2|0_i\rangle = |0_i\rangle \qquad (\hat{\sigma}_i^z)^2|1_i\rangle = |1_i\rangle \tag{3.26}$$

Considering the above equations we rewrite the evolution operator in the interaction picture as

$$\hat{U}'(t) = exp[-i\lambda t\Sigma_{j>i}(I + \hat{\sigma}_i^z\hat{\sigma}_j^z)] \tag{3.27}$$

Now we perform unitary transformation to derive the evolution operator of the system [23]

$$\hat{U}(t) = e^{-i\hat{H}_0 t}\hat{U}'(t) = exp[-it(\eta/2)\Sigma_i\hat{\sigma}_i^z - i\lambda t\Sigma_{j>i}(I + \hat{\sigma}_i^z\hat{\sigma}_j^z)] \tag{3.28}$$

In order to obtain an *Ising-like Operator* we need to impose the following condition on Rabi frequency of the oscillating electric field [23]

$$\eta/2 = (N-1)\lambda \implies \hat{U}(t) = exp[-iN\lambda t\Sigma_i\hat{\sigma}_i^z + i\lambda t\Sigma_i\hat{\sigma}_i^z - i\lambda t\Sigma_{j>i}(I + \hat{\sigma}_i^z\hat{\sigma}_j^z)] \tag{3.29}$$

where *N* is an integer. Ignoring the high frequency terms in the evolution operator gives

$$\hat{U}(t) = exp\left(-4i\lambda t\Sigma_{j>i}\frac{I-\hat{\sigma}_i^z}{2}\frac{I-\hat{\sigma}_j^z}{2}\right) \tag{3.30}$$

This is an Ising-like Operator, since we prepared the initial states at $|-_i\rangle$, the cluster states are generated by applying the C-Phase gate between the qubits. we will apply the evolution operator on the qubit states to find the necessary condition to realize a C-Phase gate

$$\hat{U}(t)|0_i\rangle|0_j\rangle = |0_i\rangle|0_j\rangle \qquad \hat{U}(t)|0_i\rangle|1_j\rangle = |0_i\rangle|1_j\rangle$$

$$\hat{U}(t)|1_i\rangle|0_j\rangle = |1_i\rangle|0_j\rangle \qquad \hat{U}(t)|1_i\rangle|1_j\rangle = exp(-4i\lambda t)|1_i\rangle|1_j\rangle \tag{3.31}$$

Therefore if $4\lambda t = (2n+1)\pi$ then we will realize a C-Phase gate and subsequently a 1D array of double QDs can be evolved to the cluster state [23].

Since we described the qubit-qubit interaction by an Ising-like Hamiltonian (3.30) for a 1D array of qubits the cluster states (highly entangled states) can be defined as [25]

$$|\psi_N\rangle = (1/2^{\frac{N}{2}}) \otimes_{i=1}^{N} (|0_i\rangle + |1_i\rangle\hat{\sigma}_{i+1}^z) \tag{3.32}$$

Here $\otimes$ represents the tensor product between states.

## 3.4 Summary

In summary, we have theoretically demonstrated that the concept of a geometric phase gate can be generalized to circuit QED in order to generate cluster states. We could achieve the qubit entanglement by capacitive coupling of the double QDs to a TLR at avoided crossing where the coupling coefficient was maximal. We ignored the responsible term for ac-Stark shift in the interaction Hamiltonian by imposing some condition on the Rabi frequency of the classical field.

In order to perform the C-Phase gate it was necessary to apply an oscillating voltage for the right amount time. The evolution operator for this system was an Ising-like Operator. Therefore we can generate cluster states from a 1D array of double QDs.

The cluster states were created in one step by applying the oscillating voltage simultaneously to all qubits. This is one of the important features of our scheme to realize a scalable quantum computer. Also we exploited the long spin relaxation time of the double QDs for single qubit readout process by mapping the charge states to spin single states. In the next chapter we will represent a proposal to perform the single qubit readout based on the ac-Stark shift effect.

# 4. The cluster state quantum computation

In this chapter we will present a proposal to perform the single-qubit readout in circuit QED. This scheme is based on the ac-Stark shift effect that has already been demonstrated [26]. Since the transition energy of the charge qubit at the avoided crossing is almost 10μev, we are required to perform the experiment at temperatures less than 100mK [23].

## 4.1 The single-qubit readout in circuit QED

As was mentioned earlier, to prolong the spin relaxation time we will need to apply a large static magnetic field in the plane of the 2DEG [14]. Also it is important to know that the qubits must be manipulated simultaneously. That means all the signals must be in phase at the location of the double QDs. In order to prepare qubits in the ground state of the double QDs we first create the electron depletion region under the gates by applying a large negative static voltage to the gates so that no electrons are left in the dots. Then by controlling the voltage difference between the gates (detuning) we provide sufficient electrostatic energy to fill up the right QDs with two electrons (large positive detuning) figure 4.1(a). According to Pauli's exclusion principle, two electrons with the same spin cannot occupy the same energy level; therefore the lowest energy level in the doubly occupied QD will be spin singlet state $|(0,2)S\rangle$.

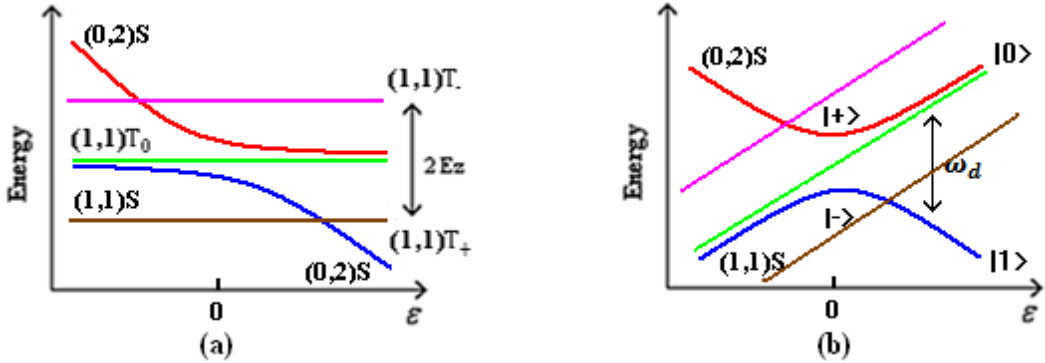

*Figure 4.1: (a) Energy diagram of a double QD in presence of an applied magnetic field. Due to the interaction between the spin of the electron and the applied magnetic field, the spin triplet state (1,1)T will split into three triplet spin states separated by the Zeeman energy, $E_z$. (b) Energy diagram in the rotating frame of the charge basis states where the charge states are mapped to spin singlet states $|0\rangle$ and $|1\rangle$ for readout.*

The next step is to transfer qubit states to the operation point (avoided crossing); we can achieve this by applying a pulse to the left gates. The passage from the spin singlet state $|(0,2)S\rangle$ to $|-\rangle$ must be fast with respect to the energy of the triplet state $|(1,1)T_+\rangle$, but adiabatic with respect to the tunneling energy [5]. We can perform such rapid adiabatic passage to avoided crossing by controlling the pulse's rise time [24].

Once we prepared the qubits in the initial states $|-_i\rangle$ an oscillating voltage is applied to the left gates of the double QDs. The frequency of the oscillating field is adjusted to be in resonance with the transition frequency of the charge qubits at the avoided crossing. After applying the oscillating voltage for the correct amount of time we will realize the C-Phase gate, where the qubits will be maximally entangled. Therefore a 1D array of double QDs can be prepared in a cluster state.

In order to perform the single qubit readout, we map the charge states to spin singlet states $|0\rangle$ and $|1\rangle$ (figure 4.1(b)). The long spin relaxation time enables us to perform the readout process for a large array of qubits. These two states can represent a two level system, so we can write the total Hamiltonian as (with $\hbar = 1$) [8]

$$\hat{H}_{total} = \omega(1/2 + \hat{a}^\dagger \hat{a}) + (\omega_d/2)\Sigma_i \hat{\sigma}_i^z + g_{eff} \Sigma_i (\hat{a}^\dagger \hat{\sigma}_i^- + \hat{a}\hat{\sigma}_i^+) \tag{4.1}$$

$$\hat{\sigma}_i^z = \hat{\sigma}_i^- \hat{\sigma}_i^+ - \hat{\sigma}_i^+ \hat{\sigma}_i^- \qquad \hat{\sigma}_i^- = |0_i\rangle\langle 1_i| \qquad \hat{\sigma}_i^+ = |1_i\rangle\langle 0_i| \qquad g_{eff} = 2g_0 T/\omega_d \tag{4.2}$$

where $g_0$ is the coupling coefficient at the avoided crossing (3.10), $\omega$ is the frequency of the fundamental mode of the TLR, 2T is the tunneling energy, $\omega_d$ is the qubit transition energy and $g_{eff}$ is the effective coupling coefficient at any adiabatic angle. The value of the effective coupling coefficient can reach zero at very large positive detuning where we will not be able to achieve the readout process as the coupling strength will be almost zero. The optimum value of the effective coupling coefficient can be determined experimentally by adjusting the detuning.

In order to achieve the QND measurement we must perform the qubit readout in dispersive mode [26]. Defining the qubit-TLR detuning as $\delta = \omega - \omega_d$, we consider the large detuning regime $\delta \gg g_{eff}$ where we perform the unitary transformation by $\hat{U}$ to derive the effective Hamiltonian [8]

$$\hat{U} = exp[(g_{eff}/\delta)(\hat{a}\hat{\sigma}_i^+ - \hat{a}^\dagger \hat{\sigma}_i^-)] \qquad \hat{H}_{eff} = \hat{U}\hat{H}_{int}\hat{U}^\dagger \tag{4.3}$$

Taylor expanding the unitary operator to second order in $g_{eff}$ and ignoring the non-energy conserving terms we find the effective Hamiltonian to be [26]

$$\hat{H}_{eff} = \Sigma_i \left(\omega + \frac{g_{eff}^2}{\delta}\hat{\sigma}_i^z\right)\hat{a}^\dagger \hat{a} + \frac{1}{2}\Sigma_i \left(\omega_d + \frac{g_{eff}^2}{\delta}\right)\hat{\sigma}_i^z \tag{4.4}$$

As ±1 are the eigenvalues of $\hat{\sigma}_i^z$, the first term represents the state dependent ac-stark shift that means the cavity mode (TLR mode) has stark shifted by $\pm g^2_{eff}/\delta$ depending on whether qubit is in the state $|0\rangle$ or $|1\rangle$ respectively. Whereas the transition frequency of the qubits has stark shifted by $n g^2_{eff}/\delta$ plus *Lamb* shifted by $g^2_{eff}/(2\delta)$ that $n$ is the number of the photons provided by the TLR [8].

We can clearly exploit this phenomenon to achieve the single qubit readout by sending a probe pulse through the TLR (figure 4.2) where by measuring the shift in frequency of the pulse we can determine that the qubit is in the state $|0\rangle$ or $|1\rangle$.

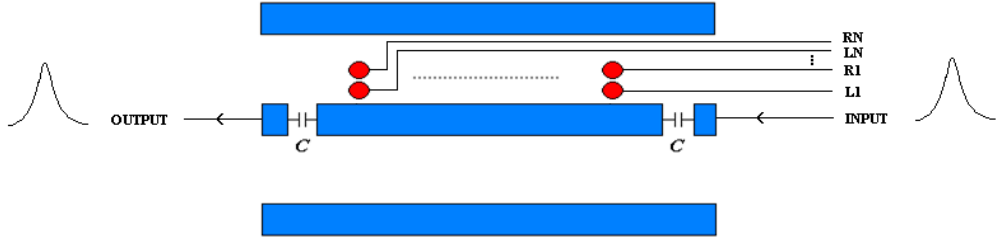

*Figure 4.2: Schematic of a TLR in circuit QED (blue), readout pulse is applied to the center microstrip line where the side ones are the grounds. C denotes the capacitive coupling of the TLR to the measuring devices. The double QDs (red circles) are capacitively coupled to the TLR and they are connected to the outside equipment via thin microstrip lines.*

In order to perform the qubit readout we send a pulse through the TLR as the reference pulse where the frequency is ac-Stark shifted, this frequency shift can be set to zero as the reference. Then we apply a pulse to left gate of the target qubit where by controlling the rise time of the pulse we can perform the rapid adiabatic passage to either $|0\rangle$ or $|1\rangle$ depending on the state of the target qubit at avoided crossing. Once the mapping process is accomplished we send the second pulse through the TLR where frequency of the TLR will be ac-Stark shifted. Let's assume the target qubit is in state $|0\rangle$ hence frequency of the pulse will be shifted by $g^2_{eff}/\delta$ (figure 4.3). This frequency shift can be stored in a classical computer as a logical bit (0 by convention).

Once the information is stored we can again set the frequency shift to zero and treat the second pulse as the reference pulse. So in order to measure the state of the second qubit, we repeat the mapping process then we send the third pulse through the TLR. The third pulse will be ac-Stark shifted by qubit one then another shift by qubit two if the frequency is shifted by $-g^2_{eff}/\delta$ it means that the qubit two is in state $|1\rangle$ and we can store this shift as the logical bit 1. We can continue this process for rest of the qubits and store the logical bits where at the end of the readout process these results can be demonstrated as an array of (0 & 1). This process can be defined as measurement based QC.

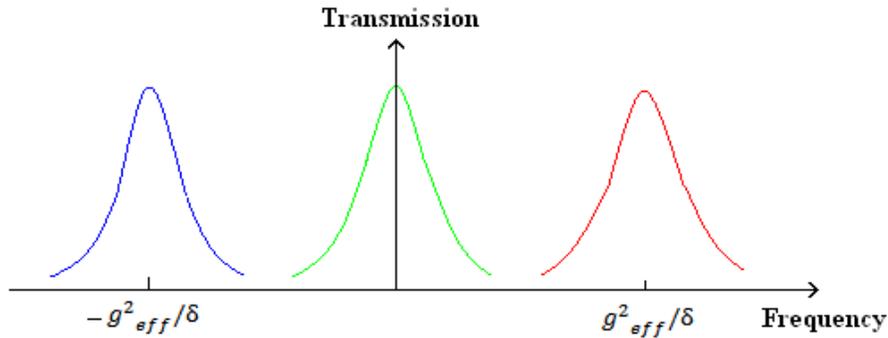

*Figure 4.3: Schematic of the ac-Stark shift experienced by the probe frequency (green). The frequency shift caused by the qubit at state $|0\rangle$ (red) or state $|1\rangle$ (blue).*

# 5. Conclusion

In this project we generalized the concept of the geometric phase gate to realize the C-Phase gate in cavity QED and circuit QED. This was essential to create the entanglement necessary to generate the cluster state. We employed a TLR to act as a cavity, which mediated the coupling between the qubits to establish the required entanglement. Since double QDs are two-level systems for the purpose of QC, we can consider the TLR as a single mode cavity as we were required to achieve the strong coupling regime.

The thermal fluctuations of the cavity would cause the phase randomization of the qubit states in the dispersive coupling regime. Since there were no other modes available for the qubits to interact with, in order to control the phase randomization we provided a coherent state to perform the Rabi oscillation on the qubits. Such coherent state with classical description would replace the quantized cavity mode where we provided this state by applying a classical field to the qubits.

The classical field would cause the ac-Stark shift that randomizes the phase of the qubit states. In order to eliminate the terms responsible for the ac-Stark shift from the interaction Hamiltonian we were required to satisfy the appropriate condition on the Rabi frequency of the qubits. This condition is satisfied as we are able to control the Rabi frequency by controlling the intensity of the classical field. Therefore, by satisfying this condition including another condition on the operation time, we realized a C-Phase gate which indicted that we have achieved an entangled state. We theoretically demonstrated that the evolution operator is an Ising-like operator and consequently the one dimensional array of the interacting qubits evolves to the cluster state.

We theoretically demonstrated that the cluster state is generated in one step by simultaneous application of an oscillating voltage to the qubits. Considering this method, including the compatibility of our proposal with the current technology of the semiconductor nanofabrication to integrate our structure on chip, makes our scheme a potential candidate for a scalable quantum computer.

The quantum computer experts have recently paid more attention to the cluster state QC as a promising candidate to build a scalable quantum computer. We proposed a scheme to perform the single-qubit readout as it is an essential part of the cluster state QC. As we mentioned in chapter four, the effective coupling at the far positive detuning will be small. This requires the experimental investigation to identify the suitable adiabatic angle for encoding the charge qubits to the spin qubits for single-qubit measurement. The outcome of this experiment can facilitate the way for further development of this technique to build the cluster states with higher dimensional arrays of the qubits.

The experimental investigation on the theory of cluster state QC requires building the cluster states with higher dimensions. The scalability of our technique can provide this opportunity to create the cluster states with higher dimensions. This is desirable because it has been proposed that the full power of quantum computation requires a higher dimensional cluster state. Studying the cluster states as the source of entanglement can reveal other features of this fascinating quantum mechanical effect.

# Appendix A

In order to derive the voltage operator for the TLR we start by writing its Lagrangian as [7][11]

$$\mathcal{L} = \int_{L/2}^{L/2} \left[ \frac{l}{2} \left( \frac{\partial Q}{\partial t} \right)^2 - \frac{1}{2C_0} \left( \frac{\partial Q}{\partial x} \right)^2 \right] dx \tag{A.1}$$

where Q(x,t) is the charge on the TLR at any instance of the time and L is the length of the TLR. We introduce K as a quantity that takes on a stationary value

$$K = \int_{t1}^{t2} \int_{L/2}^{L/2} \mathcal{L}\left(Q, , \frac{\partial Q}{\partial t}, \frac{\partial Q}{\partial x}, t\right) dx\, dt \tag{A.2}$$

we find Q(x,t) for which K is stationary

$$\delta \int_{t1}^{t2} \mathcal{L}\, dt = \alpha \left[ \frac{\partial K}{\partial \alpha} \right]_{\alpha=0} = 0 \tag{A.3}$$

$$Q(x,t,\alpha) = Q(x,t,0) + \alpha f(x,t) \tag{A.4}$$

Q(x,t,α) is unknown function for which (A.3) is satisfied and f(x,t) is an arbitrary deviation that describes the varied function Q(x,t,α). This deviation is required to be differentiable and to vanish at the end points

$$\frac{\partial Q(x,t,\alpha)}{\partial x} = \frac{\partial Q(x,t,0)}{\partial x} + \alpha \frac{\partial f(x,t)}{\partial x} \tag{A.5}$$

$$\frac{\partial Q(x,t,\alpha)}{\partial t} = \frac{\partial Q(x,t,0)}{\partial t} + \alpha \frac{\partial f(x,t)}{\partial t} \tag{A.6}$$

using (A.2)-(A.6) we find

$$\int_{t1}^{t2} \int_{\frac{L}{2}}^{\frac{L}{2}} \left( \frac{\partial \mathcal{L}}{\partial Q} - \frac{\partial}{\partial x} \frac{\partial \mathcal{L}}{\partial Q_x} - \frac{\partial}{\partial t} \frac{\partial \mathcal{L}}{\partial Q_t} \right) f(x,t)\, dx\, dt = 0 \tag{A.7}$$

since f(x,t) is arbitrary therefore we can write

$$\frac{\partial \mathcal{L}}{\partial Q} - \frac{\partial}{\partial x} \frac{\partial \mathcal{L}}{\partial Q_x} - \frac{\partial}{\partial t} \frac{\partial \mathcal{L}}{\partial Q_t} = 0,\ \frac{\partial \mathcal{L}}{\partial Q} = 0 \tag{A.8}$$

from (A.8) we extract the Euler-Lagrange equation that is describing a wave with speed of u

$$\frac{\partial^2 Q}{\partial x^2} - \frac{1}{u^2}\frac{\partial^2 Q}{\partial t^2} = 0 \quad u = \frac{1}{\sqrt{lc_0}} \tag{A.9}$$

considering charge conservation as boundary condition we solve the wave equation to find Q(x,t)

$$Q(-\tfrac{L}{2},t) = Q(\tfrac{L}{2},t) = 0 \quad Q(x,t) = \sqrt{\tfrac{2}{L}}\sum_{n_o} F_{n_o}(t)\cos(\tfrac{n_o \pi x}{L}) + \sqrt{\tfrac{2}{L}}\sum_{n_e} F_{n_e}(t)\sin(\tfrac{n_e \pi x}{L}) \tag{A.10}$$

where $n_o$ and $n_e$ are the odd and the even modes of the cavity respectively F(t) is the temporal part of the charge variation. Replacing Q(x,t) in (A.1) we obtain the Lagrangian as

$$\mathcal{L} = \sum_n \tfrac{l}{2}\dot{F}_n^2(t) - \tfrac{1}{2c_0}(\tfrac{n\pi}{L})^2 F_n^2(t) \tag{A.11}$$

Providing $\hat{a}_n^\dagger$ and $\hat{a}_n$ as the creation and the annihilation operators respectively including $P_n(t)$ the canonically conjugated momentum of $F_n(t)$ and $\omega_n$ as the nth fundamental mode frequency of the TLR we obtain

$$\hat{F}_n(t) = \sqrt{\tfrac{\hbar \omega_n c_0}{2}} \tfrac{L}{n\pi} [\hat{a}_n(t) + \hat{a}_n^\dagger(t)] \tag{A.12}$$

$$\hat{P}_n(t) = -i\sqrt{\tfrac{\hbar \omega_n l}{2}} \tfrac{L}{n\pi} [\hat{a}_n(t) - \hat{a}_n^\dagger(t)] \tag{A.13}$$

replacing (A.12) and (A.13) in the (A.10) we derive the voltage operator to be

$$\hat{V}(x,t) = \tfrac{1}{c_0}\tfrac{\partial Q(x,t)}{\partial x} = -\sum_{n_o}\sqrt{\tfrac{\hbar \omega_{n_o}}{Lc_0}}\sin(\tfrac{n_o \pi x}{L})[\hat{a}_{n_o}(t) + \hat{a}_{n_o}^\dagger(t)] +$$

$$\sum_{n_e}\sqrt{\tfrac{\hbar \omega_{n_e}}{Lc_0}}\cos(\tfrac{n_e \pi x}{L})[\hat{a}_{n_e}(t) + \hat{a}_{n_e}^\dagger(t) \tag{A.14}$$

Due to finite length of the TLR it will perform as a resonator with the resonance frequency $\omega_0 = \frac{2\pi}{L\sqrt{lc_0}}$. The TLR is coupled to the outside world by capacitance C; therefore we will need to renormalize the resonance frequency as $\omega \approx \omega_0(1-2\varepsilon_0)$ where $\varepsilon_0 = \frac{C}{Lc_0}$ [12]. Considering the first even fundamental mode frequency of the TLR we can write the voltage operator as

$$\hat{V}_{TLR}(x) = \sqrt{\tfrac{\hbar \omega}{Lc_0}}\cos(\tfrac{\pi x}{L} + \varphi)[\hat{a} + \hat{a}^\dagger] \tag{A.15}$$

where φ is the phase shift due to the frequency renormalization and satisfies $\tan\varphi = 2\pi\varepsilon_0$ [12].

# Appendix B

The free Hamiltonian describing a single mode cavity and the two identical two level atoms is represented as [18]

$$\widehat{H}_{free} = \widehat{H}_{atom} + \widehat{H}_{cavity} = \omega_0 \widehat{S}_z + \omega_a \hat{a}^\dagger \hat{a} \qquad \hbar = 1 \tag{B.1}$$

$$\widehat{S}_z = (1/2) \Sigma_{j=1,2}(|e_j\rangle\langle e_j| - |g_j\rangle\langle g_j|) \tag{B.2}$$

$\omega_0$ is the transition frequency of the atom, $\omega_a$ is the frequency of cavity mode and the atoms have been labeled by the index j.

By using the JC Hamiltonian and the rotating wave approximation the simultaneous interaction between two atoms and a single mode cavity in the presence of a class classical field can be described as [18]

$$\widehat{H}_i = \widehat{H}_Q + \widehat{H}_C = (g/2)\Sigma_{j=1,2}(\hat{a}^\dagger \widehat{S}_j^- + \hat{a}\widehat{S}_j^+) + (\Omega/2)\Sigma_{j=1,2}\left(\widehat{S}_j^+ e^{-i\omega t} + \widehat{S}_j^- e^{i\omega t}\right) \tag{B.3}$$

$$\widehat{S}_j^+ = |e_j\rangle\langle g_j| \qquad \widehat{S}_j^- = |g_j\rangle\langle e_j| \tag{B.4}$$

$\widehat{H}_i$ is the interaction Hamiltonian, $\widehat{H}_Q$ describes simultaneous interaction of the two atoms with a quantized cavity mode and we denote the atom-cavity coupling constant by g; the interaction of the two atoms with a classical field is expressed by $\widehat{H}_C$ that $\Omega$ is the Rabi frequency and $\omega$ is the frequency of the classical field. Hence the total Hamiltonian for such structure will be [18]

$$\widehat{H}_{total} = \widehat{H}_i + \widehat{H}_{free} \tag{B.5}$$

We find the interaction Hamiltonian in the interaction picture by performing the unitary transformation where we employ the Campbell-Baker-Hausdorff formula to proceed as [18]

$$\widehat{H}_{int} = e^{i\widehat{H}_{free} t} \widehat{H}_i e^{-i\widehat{H}_{free} t} = \widehat{H}_i + it[\widehat{H}_{free}, \widehat{H}_i] - t^2/2[\widehat{H}_{free}, [\widehat{H}_{free}, \widehat{H}_i]] + \cdots \tag{B.6}$$

$$[\hat{a}, \hat{a}^\dagger] = \mathbf{I} \qquad [\hat{a}^\dagger \hat{a}, \hat{a}] = -\hat{a} \qquad [\hat{a}^\dagger \hat{a}, \hat{a}^\dagger] = \hat{a}^\dagger \qquad [\hat{a}, \widehat{S}_j^\pm] = [\hat{a}^\dagger, \widehat{S}_j^\pm] = 0$$

$$[\widehat{S}_z, \widehat{S}_j^-] = -\widehat{S}_j^- \qquad [\widehat{S}_z, \widehat{S}_j^+] = \widehat{S}_j^+ \qquad [\widehat{S}_j^+, \widehat{S}_j^-] = 2\widehat{S}_z \tag{B.7}$$

$$e^{i\omega_a \hat{a}^\dagger \hat{a} t} \hat{a}^\dagger e^{-i\omega_a \hat{a}^\dagger \hat{a} t} = \hat{a}^\dagger + i\omega_a t[\hat{a}^\dagger \hat{a}, \hat{a}^\dagger] - \omega_a^2 t^2/2[\hat{a}^\dagger \hat{a}, [\hat{a}^\dagger \hat{a}, \hat{a}^\dagger]] + \cdots =$$

$$\hat{a}^\dagger + \hat{a}^\dagger i\omega_a t - \hat{a}^\dagger \omega_a^2 t^2/2 + \cdots = \hat{a}^\dagger e^{i\omega_a t} \tag{B.8}$$

$$e^{i\omega_0 \hat{S}_z t} \hat{S}_j^- e^{-i\omega_0 \hat{S}_z t} = \hat{S}_j^- + i\omega_0 t[\hat{S}_z, \hat{S}_j^-] - \omega_0^2 t^2/2\,[\hat{S}_z,[\hat{S}_z, \hat{S}_j^-]] + \cdots =$$

$$\hat{S}_j^- - \hat{S}_j^- i\omega_0 t - \hat{S}_j^- \omega_0^2 t^2/2 + \cdots = \hat{S}_j^- e^{-i\omega_0 t} \tag{B.9}$$

$$e^{i\omega_a \hat{a}^\dagger \hat{a} t} \hat{a}\, e^{-i\omega_a \hat{a}^\dagger \hat{a} t} = \hat{a} + i\omega_a t[\hat{a}^\dagger \hat{a}, \hat{a}] - \omega_a^2 t^2/2[\hat{a}^\dagger \hat{a},[\hat{a}^\dagger \hat{a}, \hat{a}]] + \cdots =$$

$$\hat{a} - \hat{a} i\omega_a t - \hat{a}\omega_a^2 t^2/2 + \cdots = \hat{a} e^{-i\omega_a t} \tag{B.10}$$

$$e^{i\omega_0 \hat{S}_z t} \hat{S}_j^+ e^{-i\omega_0 \hat{S}_z t} = \hat{S}_j^+ + i\omega_0 t[\hat{S}_z, \hat{S}_j^+] - \omega_0^2 t^2/2[\hat{S}_z,[\hat{S}_z, \hat{S}_j^+]] + \cdots =$$

$$\hat{S}_j^+ - \hat{S}_j^+ i\omega_0 t - \hat{S}_j^+ \omega_0^2 t^2/2 + \cdots = \hat{S}_j^+ e^{i\omega_0 t} \tag{B.11}$$

$$e^{i\hat{H}_{\text{free}} t} \hat{a}^\dagger \hat{S}_j^- e^{-i\hat{H}_{\text{free}} t} = e^{i\omega_a \hat{a}^\dagger \hat{a} t} \hat{a}^\dagger e^{-i\omega_a \hat{a}^\dagger \hat{a} t} e^{i\omega_0 \hat{S}_z t} \hat{S}_j^- e^{-i\omega_0 \hat{S}_z t} = \hat{a}^\dagger \hat{S}_j^- e^{-i\delta t} \tag{B.12}$$

$$e^{i\hat{H}_{\text{free}} t} \hat{a}\, \hat{S}_j^+ e^{-i\hat{H}_{\text{free}} t} = e^{i\omega_a \hat{a}^\dagger \hat{a} t} \hat{a}\, e^{-i\omega_a \hat{a}^\dagger \hat{a} t} e^{i\omega_0 \hat{S}_z t} \hat{S}_j^+ e^{-i\omega_0 \hat{S}_z t} = \hat{a}\, \hat{S}_j^+ e^{i\delta t} \tag{B.13}$$

Considering the classical field in resonance with the atoms $\omega_0 = \omega$ and $\delta = \omega_0 - \omega_a$ as the atom–cavity detuning after replacing the results of (B.8)-(B.13) in (B.6) we find the interaction Hamiltonian to be [18]

$$\hat{H}_{\text{int}} = \Sigma_{j=1,2}[(g/2)(\hat{a}^\dagger \hat{S}_j^- e^{-i\delta t} + \hat{a}\hat{S}_j^+ e^{i\delta t}) + (\Omega/2)(\hat{S}_j^+ + \hat{S}_j^-)] \tag{B.14}$$

We define the new basis states as [18]

$$|+_j\rangle = 1/\sqrt{2}(|g_j\rangle + |e_j\rangle) \qquad\qquad |-_j\rangle = 1/\sqrt{2}(|g_j\rangle - |e_j\rangle) \tag{B.15}$$

based on these basis states we rewrite the new operators as [18]

$$|+_j\rangle\langle +_j| = 1/2(|g_j\rangle\langle g_j| + |g_j\rangle\langle e_j| + |e_j\rangle\langle g_j| + |e_j\rangle\langle e_j| = 1/2(I + \hat{S}_j^+ + \hat{S}_j^-) \tag{B.16}$$

$$|-_j\rangle\langle -_j| = 1/2(|g_j\rangle\langle g_j| - |g_j\rangle\langle e_j| - |e_j\rangle\langle g_j| + |e_j\rangle\langle e_j|) = 1/2(I - \hat{S}_j^+ - \hat{S}_j^-) \tag{B.17}$$

$$\hat{\sigma}_j^z = 1/2(|+_j\rangle\langle +_j| - |-_j\rangle\langle -_j|) = 1/2(\hat{S}_j^+ + \hat{S}_j^-) \tag{B.18}$$

$$\hat{\sigma}_j^+ = |+_j\rangle\langle -_j| = 1/2(|g_j\rangle\langle g_j| - |g_j\rangle\langle e_j| + |e_j\rangle\langle g_j| - |e_j\rangle\langle e_j|) =$$

$$-\hat{S}_z - 1/2(\hat{S}_j^- - \hat{S}_j^+) \tag{B.19}$$

$$\hat{\sigma}_j^- = |-_j\rangle\langle +_j| = 1/2(|g_j\rangle\langle g_j| + |g_j\rangle\langle e_j| - |e_j\rangle\langle g_j| - |e_j\rangle\langle e_j|) =$$

$$-\hat{S}_z + 1/2(\hat{S}_j^- - \hat{S}_j^+) \tag{B.20}$$

$$\hat{S}_j^+ = \hat{\sigma}_j^z + 1/2(\hat{\sigma}_j^+ - \hat{\sigma}_j^-) \qquad\qquad \hat{S}_j^- = \hat{\sigma}_j^z + 1/2(\hat{\sigma}_j^- - \hat{\sigma}_j^+) \qquad (B.21)$$

replacing the new operators in (B.14) we obtain the interaction Hamiltonian as [18]

$$\hat{H}_{int} = \Sigma_{j=1,2}\{g/2[\hat{a}^\dagger e^{-i\delta t}(\hat{\sigma}_j^z + 1/2\hat{\sigma}_j^- - 1/2\hat{\sigma}_j^+) +$$

$$\hat{a}e^{i\delta t}(\hat{\sigma}_j^z + 1/2\hat{\sigma}_j^+ - 1/2\hat{\sigma}_j^-)] + \Omega\hat{\sigma}_j^z\} \qquad (B.22)$$

Using the Schrödinger equation to find the time evolution of the this system [18]

$$i\frac{d|\psi(t)\rangle}{dt} = \hat{H}_{int}|\psi(t)\rangle \qquad (B.23)$$

in order to find the time evolution of the system ($|\psi'(t)\rangle$) in the Schrödinger picture we perform the unitary transformation by $(\hat{H}_0 = \Omega\Sigma_{j=1,2}\hat{\sigma}_j^z)$ [18]

$$|\psi(t)\rangle = e^{-i\hat{H}_0 t}|\psi'(t)\rangle \qquad (B.24)$$

from (B.24) we can write

$$\frac{d|\psi(t)\rangle}{dt} = -i\hat{H}_0 e^{-i\hat{H}_0 t}|\psi'(t)\rangle + e^{-i\hat{H}_0 t}\frac{d|\psi'(t)\rangle}{dt} \qquad (B.25)$$

by replacing $|\psi(t)\rangle$ from (B.24) into (B.23) we will have

$$\frac{d|\psi(t)\rangle}{dt} = -i\hat{H}_{int} e^{-i\hat{H}_0 t}|\psi'(t)\rangle \qquad (B.26)$$

substituting RH side of (B.26) into LH side of (B.25) we obtain

$$\left(e^{i\hat{H}_0 t}\hat{H}_{int}e^{-i\hat{H}_0 t} - \hat{H}_0\right)|\psi'(t)\rangle = i\frac{d|\psi'(t)\rangle}{dt} \implies i\frac{d|\psi'(t)\rangle}{dt} = \hat{H}_i'|\psi'(t)\rangle \qquad (B.27)$$

where $\hat{H}_i'$ is the interaction Hamiltonian in the Schrödinger picture. Let's consider the following commutation relations between the new operators

$$[\hat{\sigma}_j^z, \hat{\sigma}_j^+] = \hat{\sigma}_j^+ \qquad\qquad [\hat{\sigma}_j^z, \hat{\sigma}_j^-] = -\hat{\sigma}_j^- \qquad (B.28)$$

substituting $\hat{H}_{int}$ from (B.23) into (B.27) and by using the Campbell-Baker-Hausdorff formula as well as the Taylor expansion we proceed to find $\hat{H}_i'$

$$e^{i\hat{H}_0 t}\hat{H}_{int}e^{-i\hat{H}_0 t} = \hat{H}_{int} + it[\hat{H}_0, \hat{H}_{int}] - t^2/2[\hat{H}_0, [\hat{H}_0, \hat{H}_{int}]] + \cdots \qquad (B.29)$$

$$e^{-i\Omega t} = 1 - it\Omega - \Omega^2 t^2/2 + \cdots \qquad (B.30)$$

$$[\hat{H}_0, \hat{H}_{int}] = \Omega g/2 \Sigma_{j=1,2} [\hat{a}^\dagger e^{-i\delta t}([\hat{\sigma}_j^z, 1/2\hat{\sigma}_j^-] - [\hat{\sigma}_j^z, 1/2\hat{\sigma}_j^+]) +$$
$$\hat{a}e^{i\delta t}([\hat{\sigma}_j^z, 1/2\hat{\sigma}_j^+] - [\hat{\sigma}_j^z, 1/2\hat{\sigma}_j^-])] \Rightarrow$$

$$[\hat{H}_0, \hat{H}_{int}] = \Omega g/2 \Sigma_{j=1,2}[\hat{a}^\dagger e^{-i\delta t}(-1/2\hat{\sigma}_j^- - 1/2\hat{\sigma}_j^+) + \hat{a}e^{i\delta t}(1/2\hat{\sigma}_j^- + 1/2\hat{\sigma}_j^+)] \quad (B.31)$$

$$[\hat{H}_0, [\hat{H}_0, \hat{H}_{int}]] = \Omega^2 g/2 \Sigma_{j=1,2} [\hat{a}^\dagger e^{-i\delta t}(-[\hat{\sigma}_j^z, 1/2\hat{\sigma}_j^-] - [\hat{\sigma}_j^z, 1/2\hat{\sigma}_j^+]) +$$
$$\hat{a}e^{i\delta t}([\hat{\sigma}_j^z, 1/2\hat{\sigma}_j^+] + [\hat{\sigma}_j^z, 1/2\hat{\sigma}_j^-])] \Rightarrow$$

$$[\hat{H}_0, [\hat{H}_0, \hat{H}_{int}]] = \Omega^2 g/2 \Sigma_{j=1,2}[\hat{a}^\dagger e^{-i\delta t}(1/2\hat{\sigma}_j^- - 1/2\hat{\sigma}_j^+) +$$
$$\hat{a}e^{i\delta t}(1/2\hat{\sigma}_j^+ - 1/2\hat{\sigma}_j^-)] \quad (B.32)$$

$$\hat{H}_i^{'} = \left(e^{i\hat{H}_0 t} \hat{H}_{int} e^{-i\hat{H}_0 t} - \hat{H}_0\right) = g/2 \Sigma_{j=1,2}\{\hat{a}^\dagger e^{-i\delta t}[\hat{\sigma}_j^z + 1/2\hat{\sigma}_j^-(1 - it\Omega - \Omega^2 t^2/2 + \cdots)$$
$$-1/2\hat{\sigma}_j^+(1 + it\Omega - \Omega^2 t^2/2 + \cdots)] + \hat{a}e^{i\delta t}[\hat{\sigma}_j^z + 1/2\hat{\sigma}_j^+(1 + it\Omega - \Omega^2 t^2/2 + \cdots) -$$
$$1/2\hat{\sigma}_j^-(1 - it\Omega - \Omega^2 t^2/2 + \cdots)]\} \Rightarrow$$

$$\hat{H}_i^{'} = g/2 \Sigma_{j=1,2}[\hat{a}^\dagger e^{-i\delta t}(\hat{\sigma}_j^z + 1/2\hat{\sigma}_j^- e^{-i\Omega t} - 1/2\hat{\sigma}_j^+ e^{i\Omega t}) +$$
$$\hat{a}e^{i\delta t}(\hat{\sigma}_j^z + 1/2\hat{\sigma}_j^+ e^{i\Omega t} - 1/2\hat{\sigma}_j^- e^{-i\Omega t})] \quad (B.33)$$

ignoring the fast oscillating terms ($\Omega \gg \delta, g$) we write the Hamiltonian as [18]

$$\hat{H}_i^{'} = g/2 \Sigma_{j=1,2}(\hat{a}^\dagger e^{-i\delta t} + \hat{a}e^{i\delta t})\hat{\sigma}_j^z = g/2(\hat{a}^\dagger e^{-i\delta t} + \hat{a}e^{i\delta t})\hat{S}_x \quad (B.34)$$

we can write the evolution operator for this Hamiltonian as [18]

$$\hat{U}^{'}(t) = e^{-iA(t)\hat{S}_x^2} e^{-iB(t)\hat{S}_x \hat{a}} e^{-iC(t)\hat{S}_x \hat{a}^\dagger} \qquad \hat{H}_i^{'} = (\hat{a}^\dagger c(t) + \hat{a}b(t))\hat{S}_x \quad (B.35)$$

In order to find the time dependent functions we can apply the Schrödinger equation [18]

$$\frac{d\hat{U}^{'}(t)}{dt} = -i\hat{H}_i^{'} \hat{U}^{'}(t) \quad \Rightarrow$$

$$\left(\frac{dA(t)}{dt}\hat{S}_x^2 + \frac{dB(t)}{dt}\hat{S}_x \hat{a}\right)\hat{U}^{'}(t) + e^{-iA(t)\hat{S}_x^2} e^{-iB(t)\hat{S}_x \hat{a}}\left(\frac{dC(t)}{dt}\hat{S}_x \hat{a}^\dagger\right) e^{-iC(t)\hat{S}_x \hat{a}^\dagger} = \hat{H}_i^{'} \hat{U}^{'}(t) \quad (B.36)$$

Now by using the Campbell-Baker-Hausdorff formula we can write

$$e^{-iB(t)\hat{S}_x\hat{a}}\left(\frac{dC(t)}{dt}\hat{S}_x\hat{a}^\dagger\right)e^{iB(t)\hat{S}_x\hat{a}}e^{-iB(t)\hat{S}_x\hat{a}}e^{-iC(t)\hat{S}_x\hat{a}^\dagger} =$$

$$\frac{dC(t)}{dt}\hat{S}_x\left(\hat{a}^\dagger - iB(t)\hat{S}_x\right)e^{-iB(t)\hat{S}_x\hat{a}}e^{-iC(t)\hat{S}_x\hat{a}^\dagger} \qquad (B.37)$$

therefore from (B.36) and (B.37) we will have

$$\left(\frac{dA(t)}{dt}\hat{S}_x^2 - iB(t)\frac{dC(t)}{dt}\hat{S}_x^2 + \frac{dB(t)}{dt}\hat{a}\hat{S}_x + \frac{dC(t)}{dt}\hat{a}^\dagger\hat{S}_x\right)\widehat{U}(t) = (\hat{a}^\dagger c(t) + \hat{a}b(t))\hat{S}_x\widehat{U}(t) \qquad =>$$

$$B(t) - B(0) = \int_0^t b(t')\,dt' = \int_0^t g/2 e^{i\delta t'}\,dt' = g/2i\delta(e^{i\delta t} - 1) \qquad (B.38)$$

$$C(t) - C(0) = \int_0^t c(t')\,dt' = \int_0^t g/2 e^{-i\delta t'}\,dt' = -g/2i\delta(e^{-i\delta t} - 1) \qquad (B.39)$$

$$A(t) - A(0) = i\int_0^t B(t')c(t')\,dt' = i\int_0^t g/2 B(t')e^{-i\delta t'}\,dt' = g^2/4\delta[t + 1/i\delta(e^{-i\delta t} - 1)] \qquad (B.40)$$

We choose the initial conditions to have A(0)=B(0)=C(0)=0 so if atom-cavity detuning satisfies $\delta t = 2\pi$ we can write the evolution operator in the interaction picture as [18]

$$\widehat{U}(t) = e^{-i\widehat{H}_0 t}\widehat{U}'(t) = e^{-i\Omega t\hat{S}_x - i\lambda t\hat{S}_x^2} \qquad \lambda = g^2/4\delta \qquad (B.41)$$

Now we apply the evolution operator on atomic states to find out whether we can realize a C-Phase gate or not, therefore we consider the following relations

$$\hat{S}_x|+_1\rangle|+_2\rangle = |+_1\rangle|+_2\rangle, \hat{S}_x|+_1\rangle|-_2\rangle = \hat{S}_x|-_1\rangle|+_2\rangle = 0, \hat{S}_x|-_1\rangle|-_2\rangle = -|-_1\rangle|-_2\rangle \qquad (B.42)$$

$$\hat{S}_x^2|+_1\rangle|+_2\rangle = |+_1\rangle|+_2\rangle, \hat{S}_x^2|+_1\rangle|-_2\rangle = \hat{S}_x^2|-_1\rangle|+_2\rangle = 0, \hat{S}_x^2|-_1\rangle|-_2\rangle = |-_1\rangle|-_2\rangle \qquad (B.43)$$

From (B.41)-(B.43) we can write

$$\widehat{U}(t)|+_1\rangle|+_2\rangle = e^{-i(\Omega+\lambda)t}|+_1\rangle|+_2\rangle \qquad (B.44)$$

$$\widehat{U}(t)|+_1\rangle|-_2\rangle = |+_1\rangle|-_2\rangle \qquad (B.45)$$

$$\widehat{U}(t)|-_1\rangle|+_2\rangle = |-_1\rangle|+_2\rangle \qquad (B.46)$$

$$\widehat{U}(t)|-_1\rangle|-_2\rangle = e^{i(\Omega-\lambda)t}|-_1\rangle|-_2\rangle \qquad (B.47)$$

According to these relations in order to realize a C-Phase gate we will need to include another conditions as

$$g = \delta => \lambda t = tg^2/4\delta = \pi/2 \qquad \Omega t = (2n + 1/2)\pi =>$$

$$(\Omega - \lambda)t = 2n\pi \qquad (\Omega + \lambda)t = (2n + 1)\pi \qquad (B.48)$$

where n is an integer and bearing in mind that we need to have $\Omega \gg \delta, g$. Based on these conditions we can realize a C-Phase gate

$$\hat{U}(t)|+_1\rangle|+_2\rangle = -|+_1\rangle|+_2\rangle \tag{B.49}$$

$$\hat{U}(t)|+_1\rangle|-_2\rangle = |+_1\rangle|-_2\rangle \tag{B.50}$$

$$\hat{U}(t)|-_1\rangle|+_2\rangle = |-_1\rangle|+_2\rangle \tag{B.51}$$

$$\hat{U}(t)|-_1\rangle|-_2\rangle = |-_1\rangle|-_2\rangle \tag{B.52}$$